\documentclass[12pt]{article}

\usepackage{latexsym,amssymb,epsfig}
\input{diagrams}

\DeclareFontFamily{OT1}{rsfs10}{}
\DeclareFontShape{OT1}{rsfs10}{m}{n}{ <-> rsfs10 }{}
\DeclareMathAlphabet{\mathscript}{OT1}{rsfs10}{m}{n}

\newcommand{\eref}[1]{(\ref{#1})}
\newcommand{\sref}[1]{\S\ref{#1}}

\newcommand{\fref}[1]{Figure~\ref{#1}}
\newcommand{\cref}[1]{Chapter~\ref{#1}}
\newcommand{\bcenter}{\begin{center}}
\newcommand{\ecenter}{\end{center}}
\newcommand{\beq}{\begin{equation}}
\newcommand{\eeq}{\end{equation}}
\newcommand{\bea}{\begin{eqnarray}}
\newcommand{\eea}{\end{eqnarray}}
\newcommand{\bean}{\begin{eqnarray*}}
\newcommand{\eean}{\end{eqnarray*}}
\newcommand{\ba}{\begin{array}}
\newcommand{\ea}{\end{array}}
\newcommand{\ben}{\begin{enumerate}}
\newcommand{\een}{\end{enumerate}}
\newcommand{\bi}{\begin{itemize}}
\newcommand{\ei}{\end{itemize}}
\newcommand{\bd}{\begin{description}}
\newcommand{\ed}{\end{description}}
\def\fnote#1#2{\begingroup\def\thefootnote{#1}\footnote{#2}
     \addtocounter{footnote}{-1}\endgroup}

\def\IC{\mathbb{C}}
\def\IE{\mathbb{E}}

\def\IR{\mathbb{R}}

\def\IZ{\mathbb{Z}}

\def\IP{\mathbb{P}}

\def\Hom{{\rm Hom}}
\def\dim{{\rm dim}}
\def\cN{{\mathcal N}}
\def\cM{{\mathcal M}}

\def\cO{{\mathcal O}}

\def\nn{\nonumber}

\def\Tr{{\mathop {\rm Tr}}}
\def\to{\rightarrow}

\newcommand{\mat}[1]{\left( \matrix{#1} \right)}

\newcommand{\scaption}[1]{\caption{\small{\sf{#1}}}}

\newarrow{DDTo}{}{dash}{}{dash}>
\def\ol{\overline}
\def\ext{{\rm Ext}}
\def\coords{co\"ordinates~}
\newcommand{\diff}[2]{\frac{\partial #1}{\partial #2}}

\newcommand{\drawsquare}[2]{\hbox{
\rule{#2pt}{#1pt}\hskip-#2pt
\rule{#1pt}{#2pt}\hskip-#1pt
\rule[#1pt]{#1pt}{#2pt}}\rule[#1pt]{#2pt}{#2pt}\hskip-#2pt
\rule{#2pt}{#1pt}}
\newcommand{\fund}{\raisebox{-.5pt}{\drawsquare{6.5}{0.4}}}
\newcommand{\antifund}{\overline{\fund}}

\newtheorem{definition}{\sf DEFINITION}

\begin{document}

\begin{titlepage}

\vspace{-2cm}

\title{
   \hfill{\normalsize  UPR-1086-T} \\[1em]
   {\LARGE Lectures on D-branes, Gauge Theories and Calabi-Yau
   Singularities}
\author{Yang-Hui He
        \fnote{~}{yanghe@physics.upenn.edu}\\[0.5cm]
   {\normalsize
        Department of Physics and Math/Physics RG} \\
   	{\normalsize University of Pennsylvania} \\
   {\normalsize Philadelphia, PA 19104--6396, USA}}
\date{}
}

\maketitle

\begin{abstract}
These lectures, given at the Chinese Academy of Sciences for the
BeiJing/HangZhou International Summer School in Mathematical Physics,
are intended to introduce, to the beginning student in string theory
and mathematical physics, aspects of the rich and beautiful subject of
D-brane gauge theories constructed from local Calabi-Yau spaces.
Topics such as 
orbifolds, toric singularities, del Pezzo surfaces as well
as chaotic duality will be covered.
\end{abstract}

\thispagestyle{empty}

\end{titlepage}

\tableofcontents

\vspace{1in}

\setcounter{equation}{0}
\section{Introduction}
``For any thing so overdone is from the purpose of playing, whose
end, both at the first and now, was and is, to hold, as
'twere, the mirror up to Nature.'' And thus The Bard has so well
summed up ({\it Hamlet, III.ii.5}) a sometime too blunted purpose of
String Theory, that she, whilst enjoying her own Beauty, should not forget
to hold her mirror up to Nature and that her purpose, as a handmaid to
Natural Philosophy, is to reflect a Greater Beauty in Nature's design. 
These
lectures, as I was asked, are intended to address an audience equally
partitioned between students of mathematics and physics. I will 
attempt to convey the little I know on some aspects of the deep and
elegant interactions between physics and mathematics within the
subject of gauge theories on D-brane world-volumes arising from
compactifications on Calabi-Yau spaces; I will try to inspire the
physicist with the astounding mathematical structures and to hearten
the mathematicians with the insightful physical computations, but I
shall always emphasise an underlying theme of Nature, that this subject
of studying gauge theories arising from string compactifications is,
{\it sicut erat in principio, et nunc, et semper,}  motivated by the
pressing need of uniquely obtaining the Standard Model from string
theory.

The lectures are entitled D-branes, gauge theories and Calabi-Yau
singularities. 
I must motivate the audience as to why we wish to study these
concepts.
First, let me address the physicists. The gauge theory aspect is
clear. Depending on the particulars, string theory possesses, {\it ab
initio}, a plethora of gauge symmetries, from
the Chan-Paton factors of the open string to the $E_8 \times E_8$ or
$Spin(32)/\IZ_2$ groups of the heterotic string. Our observable world
is a four-dimensional gauge theory with the group $SU(3) \times SU(2)
\times U(1)$ with possible but not-yet-observed supersymmetry (SUSY);
methods must be devised to reduce the gauge group of string theory
thereto.

Of course, of equal
pertinence is the need to reduce dimensionality. The ten dimensions of
string theory (or the eleven dimensions of the parent M-theory) can
only have four directions in the macroscopic scale. The traditional
approach has been compactification and this is where Calabi-Yau spaces
enter the arena. We will impose $\cN = 1$ supersymmetry in four
dimensions. This imposition is {\em completely independent} of string
theory. Many phenomenologists find the possibility of $\cN = 1$ SUSY
at the electroweak scale appealing because, amongst other virtues, it
helps to provide a natural solution to the hierarchy problem: the
amount of fine-tuning needed to make the Higgs mass at the
electro-weak scale.
The six-dimensional space for compactification
is constrained by this imposition of
low-energy supersymmetry to be a Calabi-Yau (complex) threefold.

Why, then, D-branes? As it is by now well-known, string theory
contains one of the most amazing dualities, viz., Maldacena's
AdS/CFT duality, relating bulk string theory with a holographic
boundary gauge theory. The root of this correspondence is the
open/closed string duality wherein the open strings engender the
gauge theory while the closed strings beget gravity. The boundary
conditions for the open strings are D-branes, which are dynamical
objects in their own right. Therefore, our stringy compactification
necessarily includes the subject of D-branes probing Calabi-Yau
spaces.

The paradigm we will adopt is a ``brane-world'' one. We let our world
be a slice in the ten-dimensions of the Type II superstring. In other
words, we let the four dimensional world-volume of the D3-brane carry
the requisite gauge theory, while the bulk contains gravity. Therefore,
as far as the brane is concerned, the six transverse Calabi-Yau
dimensions can be modeled as non-compact (affine) varieties. This
``compactification'' by non-compact Calabi-Yau threefolds greatly
simplifies matters for us. Indeed, an affine variety that locally
models a Calabi-Yau space is far easier to handle than the compact
manifold sewn together by local patches.

The draw-back, or rather, the boon - for a myriad of rich phenomena
germinates - is the obvious fact that the
only smooth local Calabi-Yau threefold is $\IC^3$. We are thus
inevitably lead to
the study of singular Calabi-Yau varieties. This is reminiscent of the 
fact that in standard heterotic phenomenology
it is the singular points in the moduli space of
compactifications that are of particular interest. 
The qualifiers local, non-compact, affine and singular will thus be
used interchangeably henceforth.
In summary then,
our D-brane resides transversely to a singular non-compact Calabi-Yau
threefold. On the D-brane world-volume we will live and observe
some low-energy effective theory of an $\cN=1$ extension of the
Standard Model.

Now, let me turn to address the mathematicians. Of course, the
astonishing phenomenon of mirror symmetry for Calabi-Yau threefolds
has become a favoured theme in modern geometry. Mirror pairs often
involve singular manifolds and in particular, quotients. Indeed, the
local mirror programme has been extensively used to compute
topological string amplitudes and, hence, Gromov-Witten invariants
for counting curves. Thus far, affine Calabi-Yau singularities
used in local mirrors have been predominantly toric varieties such as
the conifold and cones over del Pezzo surfaces. These, together with
above-mentioned quotient spaces, shall also constitute the primary
examples which we will study.

The resolution of singularities has been a classic and ongoing
subject. Ever since McKay's discovery of the correspondence
between the finite discrete
subgroups of $SU(2)$ and affine simply-laced Lie algebras, geometers
have been attempting to explain this correspondence using resolutions
of $\IC^2$ quotients and to extend it to higher dimensions. Now, we
have a new tool. 

String theory, being a theory of extended objects, is
well-defined on such singularities. As far back as the 1980's, Dixon,
Harvey, Vafa and Witten had realised that closed strings can propagate
unhindered on orbifolds. In addition, they made a simple prediction
for the Euler character of the orbifold in terms of the resolved space,
prompting the study by mathematicians such as V.~Batyrev,
S.~S.~Roan and Y.~B.~Ruan.

The open-string sector of the story, initiated by the investigations
of Douglas and Moore, is concerned with D-brane resolutions of
singularities. Quiver theories that arise have been used by Ito,
Nakajima, Reid, Sardo-Infirri et al.~to understand the essence of the
McKay correspondence. Recent advances, notably by Bridgeland, King and
Reid, have understood and re-casted it as an auto-equivalence in the
bounded derived category of coherent sheafs on the resolved
space. Indeed, this is closely related to the physicist's understanding of
D-branes precisely as objects in the derived category.

Realising branes as such objects, or more loosely,
as supports of vector bundles (sheafs) is the
mathematician's version of brane-worlds. With the help of the works of 
Donaldson, Uhlenbeck and Yau, wherein solving for the phenomenological
constraints of super-Yang-Mills theory in four dimensions during
compactification has been reduced to constructing polystable vector
bundles, one could move from the differential to the nominally simpler
algebraic category. Thus, gauge theory on branes are intimately
related to algebraic constructions of stable bundles.

In particular, D-brane gauge theories manifest as a natural
description of symplectic quotients and their resolutions
in geometric invariant theory. Witten's gauged
linear sigma model description, later utilised by
Aspinwall, Greene, Morrison et al. as a method of finding the vacuum
of the gauge theory, provided a novel perspective on symplectic
quotients, especially toric varieties. In summary then, our D-brane,
together with the stable vector bundle (sheaf) supported thereupon,
resolves the transverse Calabi-Yau singularity which is the vacuum for
the gauge theory on the world-volume as a GIT quotient.

\begin{figure}
\centerline{\psfig{figure=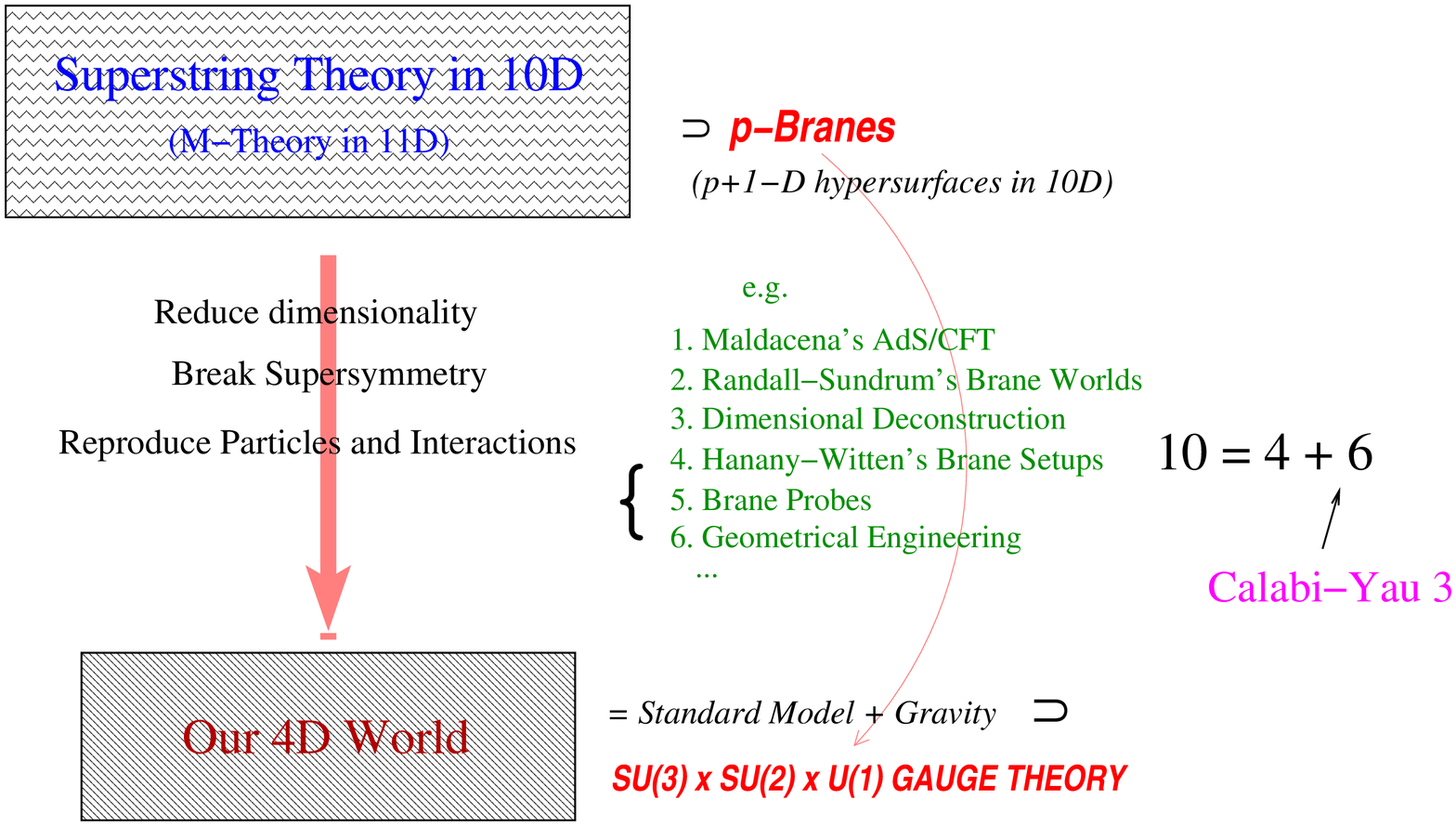,width=6in}}
\scaption{A pictorial representation of our motivations.
We need to reduce the 10 dimensions of superstring theory with various
gauge groups down to a 4 dimensional world with $\cN=1$ supersymmetry
and $SU(3) \times SU(2) \times U(1)$ gauge group with specific matter
content and interactions. Various techniques have been adopted. The
one we will study here is that of D3-branes probing a transverse
Calabi-Yau threefold singularity.
}\label{f:motivation}
\end{figure}

Hopefully, I have given ample reasons why both physicists and
mathematicians alike should study D-brane gauge theories on singular
Calabi-Yau spaces. The motivations are summarised in
\fref{f:motivation}.
Without much ado, let me proceed to the lectures.
The students can refer to \cite{thesis} wherein most of 
the material in the first two lectures are expanded.

\setcounter{equation}{0}
\section{Minute Waltz on the String}
To set the arena let me very rapidly
present to the neophytes, the necessary ingredients
from type IIB superstring theory which will be used \cite{text}.
The section is hopefully to be perused in a single minute.
The theory is a ten dimensional one with 32 super-charges and
there is a spinor generator
corresponding to the $2^{10/2} = 32$ dimensional representation of the
Clifford algebra. This is a theory of closed strings, i.e.,~mapping from
$S^1$ to the Minkowski $\IR^{1,9}$ spacetime. Bosonic particles are
excitations on the world-sheet, which is here a cylinder, traced out
by $S^1$ in spacetime. Fermionic particles also exist, as is required
by supersymmetry. Indeed, spacetime supersymmetry is induced by
the existence of worldsheet fermions the boundary conditions on which
first gave the name type IIB.

\begin{figure}[ht]
\centerline{\psfig{figure=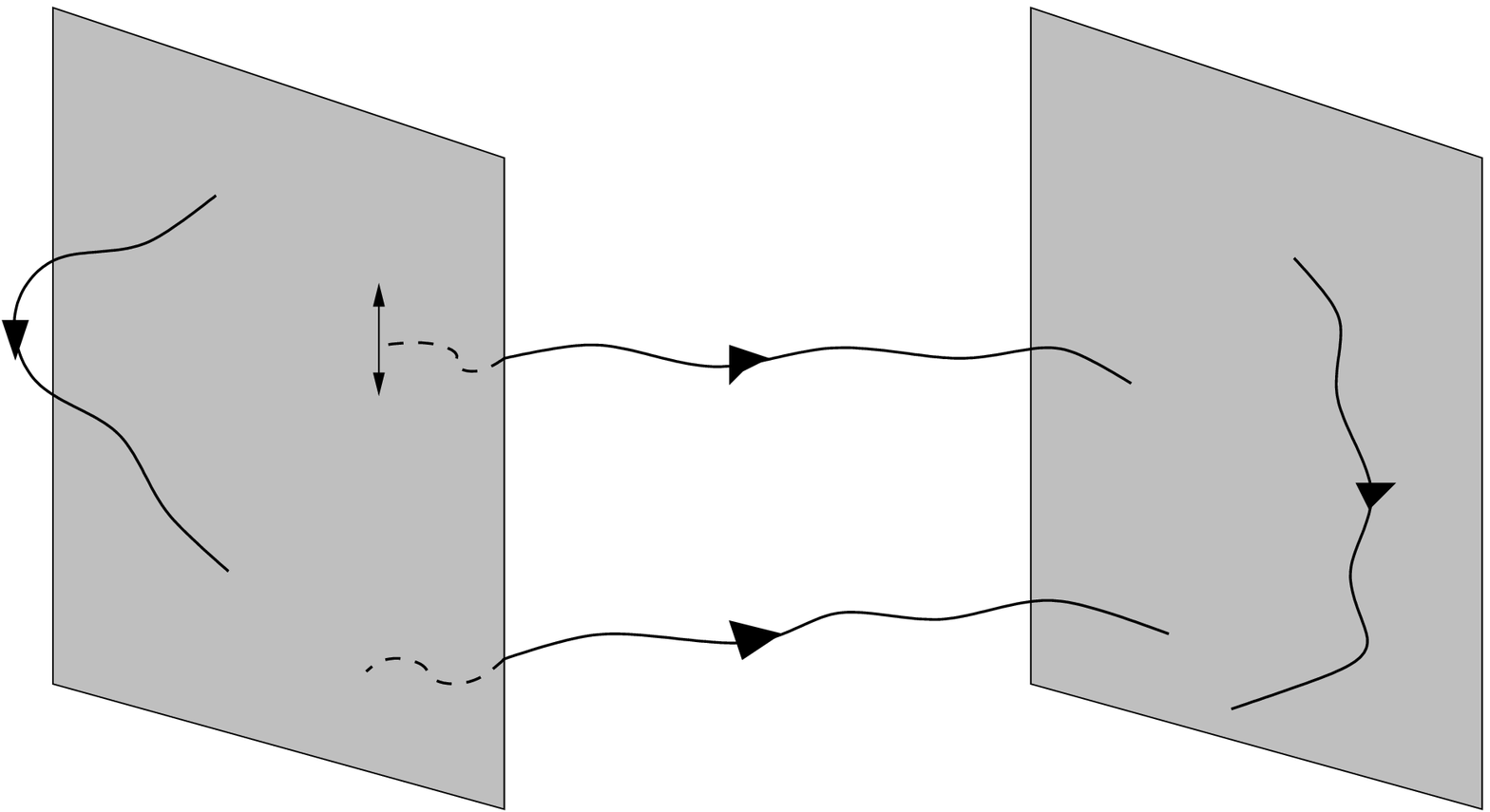,width=3.5in}}
\scaption{Open strings stretched between parallel D-branes. On the
world-volume of each brane is a $U(1)$ gauge bundle. As the two
coincide, the gauge group is enhanced to $U(2)$.
}\label{f:dbrane}
\end{figure}

By the closed/open duality inherent in string theory, that the
existence of one necessitates that of the other (as the tree-level
amplitude of the closed string is the vacuum loop of the open), we
also have open strings in the theory. They must end on subspaces of
$\IR^{1,9}$. The subspaces which provide Dirichlet boundary conditions
for the ends of the open strings are known as {\bf D-branes}. 
We shall call one with a $p+1$-dimensional world-volume a Dp-brane.
Pictorially,
this is represented in \fref{f:dbrane}. 
Polchinski's \cite{joe}
realisation, that Dp-branes are dynamical objects carrying
$p+1$-form charges, brought D-branes on an equal footing as the
fundamental string. In type IIB, $p$ will take values of all odd
integers from 1 to 10. For our purposes, we will henceforth
take $p=3$, and our
world will be $3+1$-dimensional, as it is so observed.

\subsection{The D3-brane in $\IR^{1,9}$}
Now, open strings have in their spectrum, a massless vector particle,
i.e., a $U(1)$ gauge field. Therefore, 
the D-brane must carry a $U(1)$ gauge connexion on its world-volume so
as to
accommodate the charge on the ends of the open string.
It is in this sense that we
consider the D-brane as supporting a $U(1)$ vector bundle (sheaf).
When we place a
stack of $n$ parallel D-branes coincident upon each other, we would
na\"{\i}vely expect a $U(1)^n$ gauge theory on the world
volume. Instead, we have an enhancement to the non-Abelian group
$U(n)$. 
As can be seen from \fref{f:dbrane}, this {\bf gauge enhancement}
is due to the
open strings which are stretched between the parallel branes. The
masses of these strings are proportional to the distance between the
branes and thus as the brane become coincident, we have massless
particles that are precisely the extra gauge fields.

\subsection{D3-branes on Calabi-Yau threefolds}
So far we have a ten dimensional theory of superstrings, and D-branes
on which there could be an $U(n)$ gauge group.
This, of course, is quite far from a Standard
Model in four dimensions. 
We now follow the canonical practice of considering string
theory not on $\IR^{1,9}$ but on $\IR^{1,4} \times M^{(6)}$ where 
$M^{(6)}$ is some internal manifold at the string scale, too small to
be observed. This is known as {\bf compactification}, 
an idea dating back to
T.~Kaluza and O.~Klein in 1926. As mentioned earlier, we shall require
that our $\IR^{1,3}$ universe have $\cN=1$ supersymmetry (which may
be subsequently broken at a lower energy scale). This translates to
the existence of covariantly constant spinors on $M^{(6)}$ that would
function as the supersymmetric charge.

The solution for $M^{(6)}$ is that it is
(1) compact,
(2) complex (i.e., of dimension $3 = 6 \div 2$),
(3) K\"{a}hler (the metric $g_{\mu \bar{\nu}}$ should equal to 
	$\partial_\mu \partial_{\bar{\nu}} K$ for some scalar $K$) and  
(4) has $SU(3)$ Holonomy.
E.~Calabi, an old gentleman of a distinguished bearing whose office is
two floors up from mine,
conjectured in 1954 that such manifolds should admit a
unique Ricci-flat metric in each K\"ahler class. It was only until
1971 that this conjecture was proven by S.-T.~Yau, who has been
gracious to organise this summer school, by a {\it tour de
force} differential analysis. In their honour, $M^{(6)}$ is called a
{\bf Calabi-Yau threefold}.

As far as our brane-world is concerned,
we have four dimensional D3-branes in $\IR^{1,4} \times M^{(6)}$ 
on which there
is an $U(n)$ gauge group and transverse to which gravity
propagates. We take the $\IR^{1,4}$ to be precisely the world-volume
of the brane and the transverse directions will be Calabi-Yau. In this
scenario, the Calabi-Yau manifold is to be taken as non-compact,
filling the remaining six dimensions. In other words, $M^{(6)}$ is an
affine variety that locally models a Calabi-Yau threefold. 

Indeed, if 
$M^{(6)}$ were smooth, then it can only be the trivial case of
$\IC^3$; we will address this case in the following
section. Therefore, we are lead to $M^{(6)}$ being singular. 
We will see that
it is exactly the singular structure of the geometry which aids us
phenomenologically: it will break $U(n)$ into product gauge groups, it
will reduce supersymmetry and it will yield particles that transform
under the gauge factors. We remark that the general problem of
D-branes on compact Calabi-Yau manifolds, instead of a mere affine
patch, is an extremely difficult one and excellent reviews may be
found in \cite{derived,Aspinwall}. One reason why we choose the brane-world
paradigm wherein the transverse space can be taken as a non-compact
local model, is that technically this is a much simpler problem.

A point, almost trivial, which I must emphasise, is that, as far as the
transverse singularity is concerned, the D3-brane is a point.
This obvious fact places a crucial relationship between the D3-brane
world-volume theory and the Calabi-Yau singularity: that the latter
should parametrise the former. That is, the classical vacuum of the
gauge theory on the D3-brane should be, in explicit co\"ordinates, the
defining equation of $M^{(6)}$. I will re-iterate this point later.

\begin{figure}[h]
\centerline{\psfig{figure=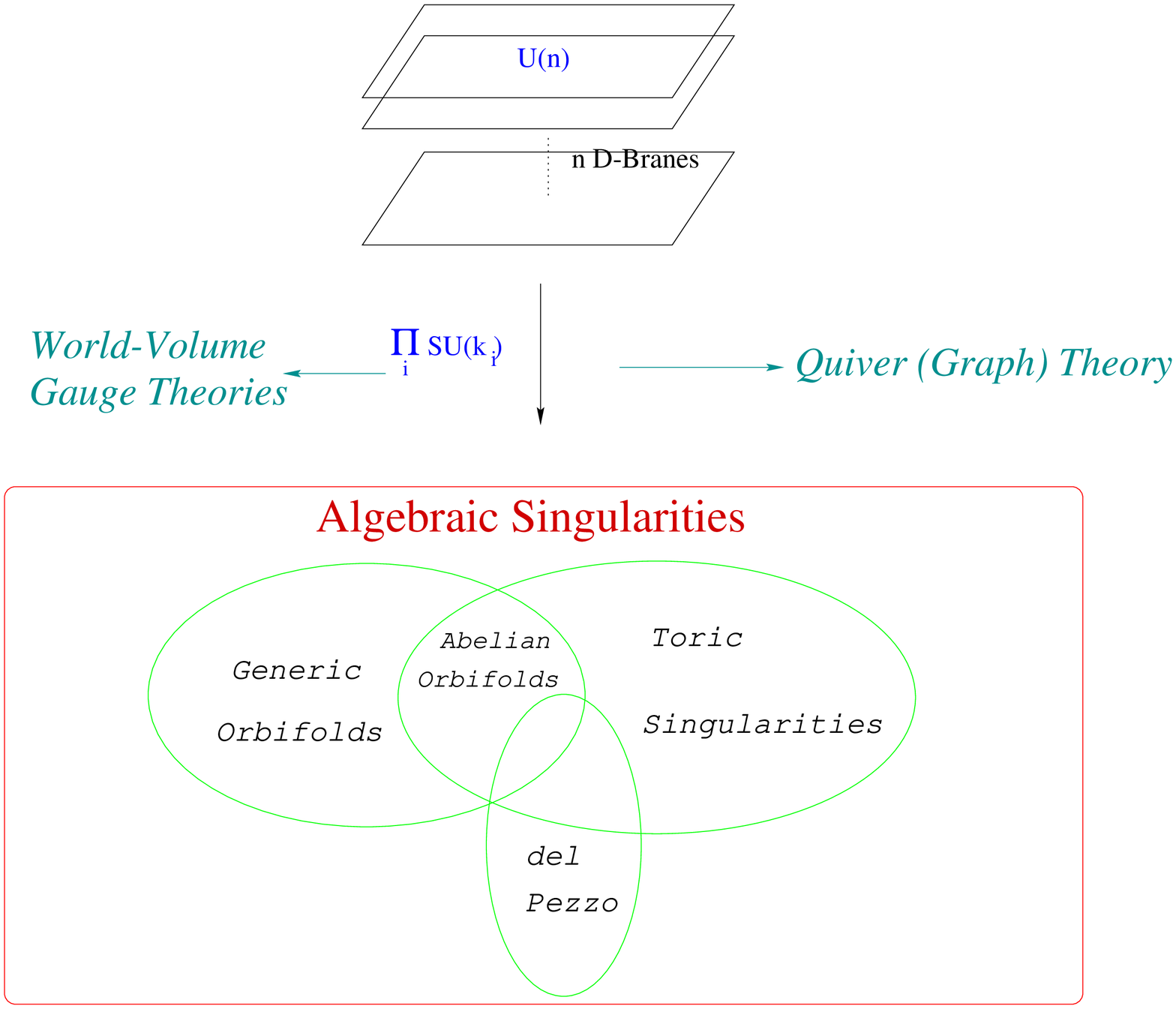,width=5in}}
\scaption{Our paradigm is to place a stack of $n$ parallel coincident
D3-branes on an affine Calabi-Yau threefold singularity $S$. The
geometry of $S$ will project the $U(n)$ gauge group on the branes to
product gauge groups with bi-fundamental matter and interactions. The
resulting theory is conveniently represented as a quiver
diagram. Examples of $S$ thus far investigated have been orbifolds,
toric singularities and cones over del Pezzo surfaces, as shewn in the
Venn diagram.
}\label{f:state}
\end{figure}

For now, let us summarise the philosophy pictorially in \fref{f:state}.
We place a stack of $n$ D3-branes transverse to a Calabi-Yau threefold 
$M^{(6)}$, which, since it is singular, we will henceforth call $S$.
The local model $S$ will afford some explicit description as an affine
variety. The geometry of $S$ will project the $U(n)$ gauge theory to
product gauge groups (ultimately that of the Standard Model).
The singularities so far used have been orbifolds, toric singularities
and cones over del Pezzo surfaces, the relations amongst which are
drawn in the Venn diagram in the figure.

\setcounter{equation}{0}
\section{The Simplest Case: $S = \IC^3$}
Let me begin with the simplest non-compact 
Calabi-Yau threefold. This is, of course, when $S$ is $\IC^3$, which
is trivially Ricci flat. Here, the D3-brane freely
propagates in flat space. The world volume theory has a $U(n)$ gauge
group as mentioned above. The presence of the brane breaks the
$SO(1,9)$ Lorentz symmetry of $\IR^{1,9}$, whereby breaking half of
the supersymmetry and we are left with 16 supercharges. In four
dimensions, this amount of supercharges corresponds to $16 / 2^{4/2} =
4$ supersymmetry. We therefore have $\cN = 4$ superconformal (SCFT) $U(n)$
gauge theory on the world-volume.

This gauge theory, is the famous boundary SCFT for
Maldacena's Correspondence \cite{juan}. Of course, the D3-brane will
warp the flat space metric to that of $AdS_5 \times S^5$ and the bulk
geometry is not strictly $\IC^3$. However, as stated above,
we are only concerned with the local gauge theory and 
not with gravitational back-reaction, therefore it suffices to
consider $S$ as $\IC^3$.

The matter content of the theory is as follows. There is a gauge field
$A^\mu$ under the $U(n)$ group. Moreover, there is an $SU(4)$
R-symmetry inherent to the $\cN=4$ SCFT which essentially is a
rotation of the four supersymmetries. Of course, in the AdS/CFT
picture, this $SU(4) \simeq SO(6)$ is the isometry group of the $S^5$
factor in $AdS_5 \times S^5$. Under this R-symmetry, there are 
Weyl fermions $\Psi_{IJ}^{\bf{4}}$, $I,J = 1, \ldots, n$, 
transforming as the ${\bf 4}$ of
the $SU(4)$ and as adjoints under $U(n)$. The SUSY partners are
bosonic fields $\Phi_{IJ}^{\bf{6}}$ under the ${\bf 6}$ of the
$SU(4)$. The superpotential is uniquely determined by the matter. In
terms of the three chiral superfields $C^i$, it is simply $W = \Tr(C^i
C^j C^k) \epsilon_{ijk}$.
For the mathematicians in the audience, we will consider $A^{\mu}$ as 
$\Hom(\IC^n, \IC^n)$, $\Psi_{IJ}^{\bf{4}}$ as 
${\bf 4} \otimes \Hom(\IC^n, \IC^n)$, and $\Phi_{IJ}^{\bf{6}}$ as
${\bf 6} \otimes \Hom(\IC^n, \IC^n)$.
What I have described above we shall call the ``parent theory.''
Her progeny will be the subject matter of these lectures.

\setcounter{equation}{0}
\section{Orbifolds and Quivers}\label{s:orb}
The next best things to $\IC^3$ are its quotients. This simple class
of singularities is called {\bf orbifolds} 
and has been studied as far
back as the 1960's by Satake et al. \cite{satake}. 
Here, I take $\IC^3$
and quotient is by some discrete finite group $\Gamma$.
The group action is
\beq
(\gamma \in \Gamma) : (x,y,z) \to \gamma \cdot (x,y,z),
\eeq
where the element $\gamma$ is written in explicit matrix representation 
and $(x,y,z)$ are the complex \coords of $\IC^3$.
We see that the origin $(0,0,0)$ is a fixed point. Because of this
fixed point, i.e., the action is not free, the quotient $\IC^3 /
\Gamma$ consisting of equivalence class under the group action, is not
a smooth manifold.

Certainly, $\Gamma$ cannot be arbitrary. In order that $S = \IC^3 /
\Gamma$ be a
Calabi-Yau singularity, it must admit a resolution to a smooth 
Calabi-Yau manifold $\tilde{S}$. This is known as a {\bf crepant
resolution}, i.e., for the map $f: \tilde{S} \to S$,
\beq
f^*K_S = K_{\tilde{S}} = \cO_{\tilde{S}},
\eeq
where $K_S$ and $K_{\tilde{S}}$ are the canonical sheaves of $S$ and
$\tilde{S}$ respectively and $K_{\tilde{S}}$ is trivial since
$\tilde{S}$ is Calabi-Yau. The subject of crepant resolutions is one
in its own right and the mathematicians in the audience are referred
to, for example, \cite{itoreid,ito}. For our purposes I will take
$\Gamma$ to be a discrete finite subgroup of $SU(3)$, i.e., the
holonomy of $\tilde{S}$. This is not a sufficient condition for
crepancy, but the techniques we introduce below work in general. 
For $\Gamma \subset SU(2)$, i.e., $S = \IC \times \IC^2 / \Gamma$,
however, they 
all have crepant resolutions and have a beautiful structure which we
will present later. The case of $\Gamma \subset (SU(4) \simeq SO(6))$ 
is also
possible \cite{su4}
but we need to consider $S = \IR^6 / \Gamma$; this is, for
now, of less interest to us because it is a real quotient and
preserves no SUSY.

%
\subsection{Projection to Daughter Theories}
As initiated in the study of Douglas and Moore \cite{DM}, with
cases addressed by Johnson and Meyers \cite{JM}, and the methodology
formalised by Lawrence, Nekrasov and Vafa \cite{LNV}, let us now study
what happens to the parent theory due to $S$. The prescription is
straight-forward: we will use elements $\gamma$ to project out any
states not invariant under $\Gamma$. That is, only $A^\mu$,
$\Psi_{IJ}^4$ and $\Phi_{IJ}^6$ that satisfy
\beq
\gamma A^\mu \gamma^{-1} = A^\mu, \quad
\Psi_{IJ}^4 = R(\gamma) \gamma \Psi^4_{IJ} \gamma^{-1}, \quad
\Phi_{IJ}^6 = R(\gamma) \gamma \Phi^6_{IJ} \gamma^{-1}
\eeq
remain in the spectrum. We have used $R(\gamma)$ for the matter fields
since there should also be an extra induced action on the R-symmetry.
Furthermore, the resulting SUSY is the commutant of $\Gamma$ in
$SU(4)$. That is, the R-symmetry left untouched by $\Gamma$ will serve
as the R-symmetry of the daughter theory.

More formally, in the notation of \cite{LNV}, let the irreducible
representation of $\Gamma$ be $\left\{ {\bf r}_{i}\right\}$ and 
decompose
\beq\label{Ni}
\IC^n \simeq \bigoplus\limits_{i} \IC^{N_{i}} {\bf r}_{i}
\eeq
for integer multiplicities $N_i$. Then, the resulting gauge group is
given by the $\Gamma$-invariant part of the gauge group
$\Hom \left( \IC^{n},\IC^{n}\right)$. That is,
\beq\label{Aproj}
\Hom \left( \IC^{n},\IC^{n}\right) ^{\Gamma}
= \bigoplus_{i,j}\left(\IC^{N_{i}}\otimes\IC^{N_{j}*}
\otimes {\bf r_i} \otimes {\bf r_j^*} \right)^\Gamma
= \bigoplus\limits_{i}\IC^{N_{i}} {\bf \otimes }
	\left( \IC^{N_{i}}\right) ^{*},
\eeq
where we have used Schur's Lemma that $({\bf r_i} \otimes {\bf
r_j^*})^\Gamma  = \delta_{ij}$ for irreducible representations 
$\left\{ {\bf r}_{i}\right\}$. In other words, the daughter gauge
group is $\prod\limits_i U(N_i)$ with $N_i$ given in \eref{Ni}. It
turns out that in the low energy effective theory the $U(1)$ factors
decouple. Therefore, in fact, the resulting gauge group is 
$\prod\limits_i SU(N_i)$.

Now, the matter fields $\Psi_{IJ}^4$ and $\Phi_{IJ}^6$ encounter a
similar projection. For ${\cal R} = {\bf 4}$ or {\bf 6}, we have
\bea\label{matterProj}
\left({\cal R} \otimes 
	\Hom \left( \IC^{n},\IC^{n}\right) \right)^{\Gamma}
&=& \bigoplus_{i,j}({\cal R} \otimes 
\left(\IC^{N_{i}}\otimes\IC^{N_{j}*}
\otimes {\bf r_i} \otimes {\bf r_j^*} \right))^\Gamma \nn \\
&=& \bigoplus_{i,j} a_{ij}^{{\cal R}} 
	\left(\IC^{N_{i}}\otimes\IC^{N_{j}*}\right),
\eea
where we have again made use of Schur's Lemma and, in addition,
the decomposition 
\beq\label{Rdecomp}
\fbox{
${\cal R}\otimes
{\bf r}_i=\bigoplus\limits_{j}a_{ij}^{{\cal R}} {\bf r}_j$.}
\eeq
In other words, the matter fields become a total of
$a_{ij}^{{\bf 4}}$ bi-fundamental fermions and 
$a_{ij}^{{\bf 6}}$ bi-fundamental bosons transforming as the
$(N_i, \bar{N}_j)$ of $SU(N_i) \times SU(N_j)$ under the product gauge
groups. To solve \eref{Rdecomp}, one uses standard orthogonality
conditions in character theory and obtain \cite{HanHe}
\beq\label{chiaij}
a_{ij}^{{\cal R}}=\frac{1}{g}\sum\limits_{\gamma
=1}^{r}r_{\gamma }\chi_{\gamma }^{{\cal R}}\chi_{\gamma }^{(i)}\chi
_{\gamma }^{(j)*},
\eeq
where $g=\left| \Gamma \right| $ is the
order of the group, $r_{\gamma }$ is the order of the conjugacy class
containing $\gamma$ and $\chi_{\gamma }^i$ is the character of
$\gamma$ in the $i$-th representation.
We summarise the daughter theories below:
\beq\label{daughter}
\begin{array}{|l|l|l|l|}
\hline
&$Parent$        & \stackrel{\Gamma}{\longrightarrow}
                        &$Orbifold Theory$\\
\hline
$SUSY$                  &{\cal N}=4     & \leadsto  &
        \begin{array}{l}
        {\cal N}=2, {\rm~for~} \Gamma\subset SU(2)     \\
        {\cal N}=1, {\rm~for~} \Gamma\subset SU(3) \\
        {\cal N}=0, {\rm~for~} \Gamma\subset \{SU(4)\simeq SO(6)\} \\
        \end{array}\\
\hline
\begin{array}{c}
        $Gauge$ \\
        $Group$
\end{array}             &U(n)           &\leadsto& \prod\limits_{i}
	        U(N_i), \qquad 
		\sum\limits_{i} N_i \dim{\bf r}_i = n\\
\hline
$Fermion$       &\Psi_{IJ}^{\bf{4}} &\leadsto   & \Psi_{f_{ij}}^{ij} \\
$Boson$         &\Phi_{IJ}^{\bf{6}} &\leadsto   & \Phi_{f_{ij}}^{ij} \qquad
			\qquad {\cal R}\otimes
			{\bf r}_i=\bigoplus\limits_{j}a_{ij}^{{\cal
			R}} {\bf r}_j\\
\hline
\end{array}
\eeq
for
$I,J=1,...,n$ and $f_{ij}=1,...,a_{ij}^{{\cal R}={\bf 4},{\bf 6}}$.

%
\subsection{Quivers}\label{s:quiver}
A convenient and visual representation of the resulting matter content
in \eref{daughter} is the so-called {\bf quiver diagram}, originating
from the German ``K\"ocher'' \cite{Gabriel}. 
The rules are simple: it is a finite
directed
graph such that each node $i$ represents a gauge factor $SU(N_i)$ and
each arrow $i \to j$, a bi-fundamental field $(N_i, \bar{N}_j)$.
The {\bf adjacency matrix} $A$ of the graph is a $k \times k$ matrix with $k$
being the number of nodes (gauge factors) and
encodes this information by having
its entry $A_{ij}$ counting the number of arrows (bi-fundamentals)
from $i$ to $j$.
My discussing finite graphs at this point is hardly a digression. We
will see next that the very usage of quiver is undoubtedly inspired by
a remarkable correspondence.

%
\subsection{The McKay Correspondence}
Let us specify the discussions in \eref{daughter} to the case of
$\cN=2$, i.e., to orbifolds of the type $\IC^3 / \Gamma \simeq
\IC \times \IC^2 / (\Gamma \subset SU(2))$. In 1884, F.~Klein, in
finding transcendental solutions to the quintic problem \cite{klein}, 
classified
the discrete finite subgroups of $SU(2)$. These are double covers of
those of $SO(3)$, which simply constitute the symmetries of the
perfectly regular shapes in $\IR^3$, viz.~the {\bf Platonic Solids}. 
The groups fall into 2 infinite series, associated to the regular
polygons, as well as 3 exceptionals, associated with the 5 regular
polyhedra: the
tetrahedron, the cube (and its dual tetrahedron) and the icosahedron
(and its dual dodecahedron). The groups are
\beq\label{su2}\ba{|c|l|c|}
\hline
\mbox{Group} & \mbox{Name} & \mbox{Order} \\ \hline
A_n \simeq \IZ_{n+1} & \mbox{Cyclic} & n+1 \\ \hline
D_n & \mbox{Binary Dihedral} & 2n \\ \hline
E_6 & \mbox{Binary Tetrahedral} & 24 \\ \hline
E_7 & \mbox{Binary Octahedral (Cubic)} & 48 \\ \hline
E_8 & \mbox{Binary Icosahedral (Dodecadedral)} & 120 \\ \hline
\ea\eeq

It was not until 1980, almost a full century later, that a remarkable
correspondence between these groups and Lie algebras were realised by
J.~McKay \cite{mckay}, yes, the very McKay also responsible for
initiating Moonshine. It has never ceased to astounded me, this
uncanny ability of his to recognise, amidst a seeming cacophony of
sounds, a single strain of melody. I first met John in Warwick I
believe, me in my younger and even more ignorant days,
him ruddy faced and a glass of wine in hand. Overcoming my
initial trepidation by a few quick bites at my own sturdy drink - my
liver being more lively at the time - I proceeded to him with
reverence. But my intimidation was unwarranted and
with paternal patience he explained at length his new conjectures
regarding modular forms, sporadic groups and exceptional algebras,
which, alas, due part to my own wanting of knowledge and part to
the fine workings of the potent liquid upon my head, I only managed a
vague glimpse, a fuller view of which only of late, fewer hair and
dryer stomach, did I acquire in more conversations with him.

What McKay realised was that one can take
the Clebsch-Gordan decomposition for
${\cal R}$, the fundamental {\bf 2} of $\Gamma \subset SU(2)$
and $\{ {\bf r}_i \}$, the irreps. That is, one can take
\beq\label{mckaydec}
{\bf 2}\otimes {\bf r}_i=\bigoplus\limits_{j}a_{ij}^{{\bf 2}}{\bf r}_j,
\eeq
and treat $a_{ij}^{{\bf 2}}$ as the adjacency matrix of some finite
graph. Then, the graphs are precisely the Dynkin diagrams of the
affine simply-laced Lie algebras:
\beq
\psfig{figure=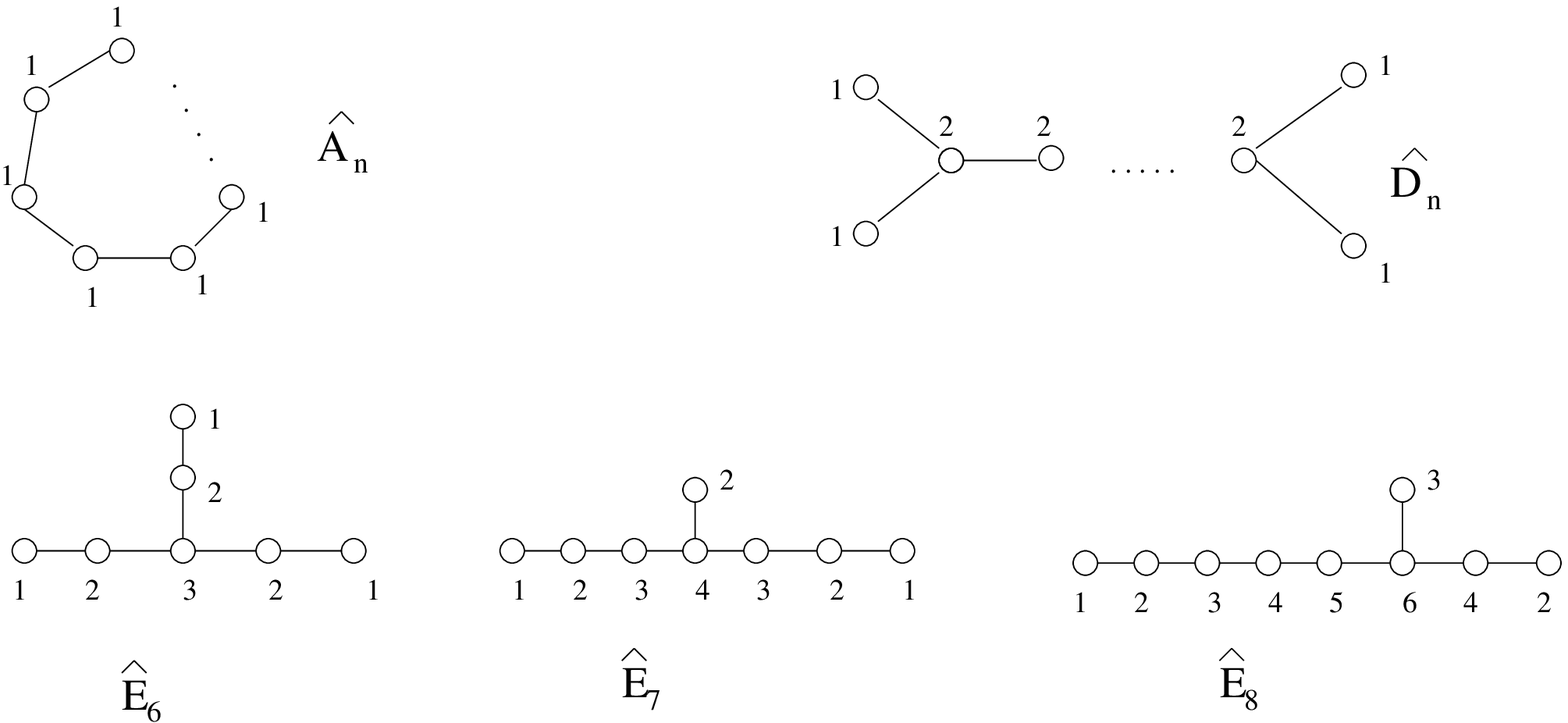,width=5in}
\eeq
In other words, the McKay quiver for
$\Gamma$ is in one-one correspondence with the affine ADE Lie algebras
(the affine diagram adds one more node to the usual Dynkin diagram,
here corresponding to the trivial representation) and the matrices
$a_{ij}^{{\bf 2}}$ in \eref{mckaydec} are precisely the Cartan
matrices of the associated algebra.
With this
hindsight, it is natural that we have so named the groups in
\eref{su2}. There are many accounts for the McKay Correspondence, and
the audience is referred to, e.g., \cite{mckaystory}.

Shortly after this discovery, algebraic geometers were busy trying to
explain this correspondence. Indeed, crepant resolution of $\IC^2 /
\Gamma$ gives the K3 surface, which, other than the trivial $T^4$, is
the only Calabi-Yau two-fold. In explicit affine \coords$\!$,  
our orbifolds are the following singularities:
\beq\label{coordade}
\begin{array}{ll}
A_n: &xy+z^n=0\\
D_n: &x^2+y^2z+z^{n-1}=0\\
E_6: &x^2+y^3+z^4=0\\
E_7: &x^2+y^3+yz^3=0\\
E_8: &x^2+y^3+z^5=0.
\end{array}
\eeq
It was soon realised by
Gonz\'alez-Springberg and Verdier \cite{GSV}, that in the crepant resolution
of \eref{coordade}, the intersection matrix of the  $-2$ exceptional
curves, i.e., the $\IP^1$-blowups, is exactly McKay's $a_{ij}^{{\bf
2}}$ in \eref{mckaydec}. Only until recently did there exist a
categorical description of the McKay correspondence in terms of an
auto-equivalence in ${\cal D}^b({\rm coh}(\widetilde{X/\Gamma})) = {\cal
D}^b({\rm coh}^G(X))$ \cite{BKR}.

\subsection{McKay, Dimension 2 and $\cN=2$}
The astute audience will recognise \eref{mckaydec} as something
earlier mentioned, viz., the matter matrix \eref{Rdecomp}.
If we take the decompositions
\bea\label{dec46}
{\bf 4} = {\bf 1}^2 \oplus {\bf 2}, \nn \\
{\bf 6} = {\bf 1}^2 \oplus {\bf 2} \oplus {\bf \bar{2}}
\eea
respectively for the fermions and bosons, then, the McKay quivers give
the matter content of the $\cN=2$ SCFT that lives on the
four-dimensional world-volume of the D3-brane which transversely
probes $\IC \times \IC^2 / (\Gamma \subset SU(2))$ (i.e., local K3).
The trivial ${\bf 1}$'s in \eref{dec46} add to nothing except diagonal
entries in \eref{chiaij}, i.e., to self-adjoining arrows at each
node. The ${\bf \bar{2}}$ in the decomposition of the {\bf 6} adds
another copy of $a_{ij}^{{\bf 2}}$ to the matter content.
One can also easily obtain the interaction terms which are nicely
presented in \cite{LNV}.

This toy model, though endowed with such a beautiful structure, is
still far from a phenomenological interest. We have $\cN=2$
supersymmetry and the theory is non-chiral, i.e., for each arrow between 
two nodes, there is another exactly in the opposite direction. 
In other words, the matter comes in conjugate pairs;
the real world, on the other hand, has chiral fermions.
Therefore, though fed a mathematical treat, we must trudge on.

%
\subsection{$\cN=1$ Theories and $\IC^3$ Orbifolds}
Phenomenologically, the most interesting case for us is $\cN=1$
theories in four dimensions. Referring to \eref{daughter}, we need
orbifolds of $\IC^3$; i.e., we need the discrete finite subgroups of
$SU(3)$. This task, was luckily performed by Blichfeldt in 1917
\cite{blich}. We summarise the classification below:
\beq\label{su3gp}
\ba{|c|c|}\hline
\mbox{Infinite Series} & \Delta(3n^2), \Delta(6n^2) \\ \hline
\mbox{Exceptionals} & 
\Sigma_{36 \times 3},
\Sigma_{60 \times 3},
\Sigma_{168 \times 3},
\Sigma_{216 \times 3}, \Sigma_{360 \times 3} \\
\hline\ea
\eeq
We see that there are two infinite series of order $3n^2$ and $6n^2$
respectively, as well as 5 exceptionals whose orders I have labelled
as subscripts. I remark that the orders are all divisible by 3, much
like the $SU(2)$ subgroups, whose orders are divisible by 2. This is
because here, in analogy with the $\IZ_2$ centre of $SU(2)$,
the centre is $\IZ_3$.

The matter content for these theories was established in
\cite{HanHe}. The fermionic ${\bf 4}$ is now ${\bf 4} = {\bf 1} \oplus
{\bf 3}$ while the bosonic ${\bf 6}$ decompose as ${\bf 3} \oplus {\bf
\ol{3}}$. The essence, then, is 
McKay-like quivers dictated by ${\bf 3}\otimes {\bf
r}_i=\bigoplus\limits_{j}a_{ij}^{{\bf 3}}{\bf r}_j$. 
We present the fermion graphs below for the exceptionals:
\beq\label{su3graph}
\psfig{figure=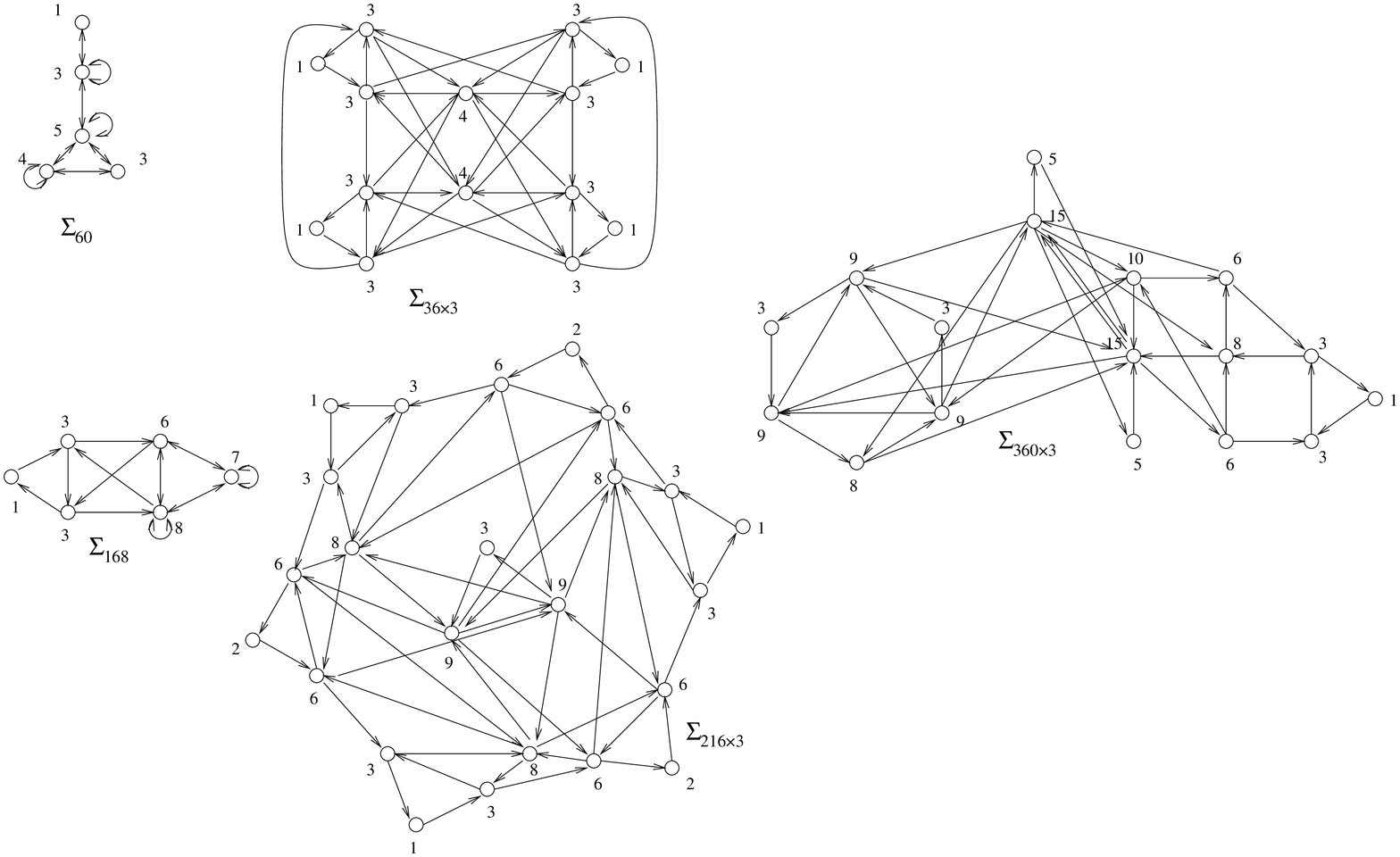,width=6in}
\eeq
We immediately see that not all arrows come with partners in the reverse
direction. This is the desired chirality for fermions. Constructing
phenomenological viable theories from these theories have been well
under way, cf.~e.g.,~\cite{KS,su3pheno}. To give a flavour of the type of
gauge groups one might obtain, I tabulate below the result for the
various subgroups. Note that I have listed more than \eref{su3gp}, by
also including the subgroups of $SU(2)$, embedded into $SU(3)$.
\beq
\begin{array}{|c|c|}
\hline
\Gamma \subset SU(3)                            &\mbox{Gauge Group} \\ \hline
\widehat{A_n} \cong \IZ_{n+1}                   &(1^{n+1}) \\
\IZ_k \times \IZ_{k'}                           &(1^{k k'}) \\
\widehat{D_n}                                   &(1^4,2^{n-3}) \\
\widehat{E_6} \cong {\cal T}                    &(1^3,2^3,3) \\
\widehat{E_7} \cong {\cal O}                    &(1^2,2^2,3^2,4) \\
\widehat{E_8} \cong {\cal I}                    &(1,2^2,3^2,4^2,5,6) \\
E_6 \cong T                                     &(1^3,3) \\
E_7 \cong O                                     &(1^2,2,3^2) \\
E_8 \cong I                                     &(1,3^2,4,5) \\
\hline\end{array}
\quad
\begin{array}{|c|c|}
\hline
\Gamma \subset SU(3)                            &\mbox{Gauge Group} \\ \hline
\Delta_{3n^2}   (n=0\mbox{ mod }3)              &(1^9,3^{\frac{n^2}{3}-1}) \\
\Delta_{3n^2}   (n\ne0\mbox{ mod }3)            &(1^3,3^{\frac{n^2-1}{3}}) \\
\Delta_{6n^2}   (n\ne0\mbox{ mod }3)            &(1^2,2,3^{2(n-1)},6^{\frac{n^2-3
n+2}{6}}) \\
\Sigma_{168}                                    &(1,3^2,6,7,8) \\
\Sigma_{216}                                    &(1^3,2^3,3,8^3) \\
\Sigma_{36\times 3}                             &(1^4,3^8,4^2) \\
\Sigma_{216\times 3}                            &(1^3,2^3,3^7,6^6,8^3,9^2) \\
\Sigma_{360\times 3}
&(1,3^4,5^2,6^2,8^2,9^3,10,15^2) 
\\
\hline\end{array}
\eeq

%
%
\subsection{Quivers, Modular Invariants, Path Algebras?}
One might ask whether as rich a structure as the abovementioned
dimension two example,
shrouded under the veil of Platonic perfection, could persist to our
present case, and, more optimistically, to higher dimension. Indeed,
the subject of various generalisations of the McKay
Correspondence is an active one, q.v.~\cite{mckay3}.

For now, I hope you could indulge me in a moment of speculation. In
\cite{HanHe}, a certain resemblance was noted between the \eref{su3graph}
and the fusion graphs of $\widehat{su(3)}$ Wess-Zumino-Witten
models, much in the same spirit of the well-known fact
that fusion graphs for $\widehat{su(2)}$
WZW models are (truncations of) ADE diagrams. Similar observations had
been independently noticed in the context of lattice models
\cite{DiFrancesco}.
Inspired by this
observation, \cite{HeSong} attempted to establish a web of
correspondences wherein stringy resolutions and world-sheet conformal
field theory are key to the McKay correspondence in dimension two and,
in specialised cases, for higher dimension. A weak but curious
relation was established in \cite{0009077} wherein the $SU(3)$ 
finite group was found to act on the terms of the modular invariant
partition function of the $\widehat{su(3)}$ WZW.

On more categorical grounds, the specialty of dimension two is even
more enforced. A theorem of P.~Gabriel dictates that the
path algebra (i.e., the algebra generated by composing arrows head to
tail) of any quiver has 
finite representation iff quiver is ADE. Therefore, 
$\IC^2$-orbifold quivers are the only finite quivers. 

The tree-level
beta function for the orbifold theories were identified with a certain
criterion for (sub)additivity of graphs in \cite{9911114}. In dimension
two, the conformality of the IR fixed point implies that the quiver
must be strictly additive, the only cases of which, by a theorem of 
Happel-Preiser-Ringel, are (generalisations of) ADE graphs. Higher
dimensional cases require an extension of the definition of
additivity, a systematic investigation
of which thus far has not been performed. It
is expected, however, that these theories are unclassifiable, a true
hindrance to the persistence of the intricate web of inter-relations
in \cite{HeSong}.

\subsection{More Games}
Before leaving the subject of orbifolds, let me entice you with a few
more games we could play. In the derivations \eref{Aproj} and
\eref{matterProj} presented above, we used the ordinary representation
of $\Gamma$. More concretely, we used explicit matrix
representations of the group elements as linear transformations. What
if we used, instead, more generalised representations such as projective
representations? These are representations $\gamma$ of $\Gamma$ such
that for any two group elements $g_{1,2} \in \Gamma$, 
\beq
\gamma(g_1) \gamma(g_2) = A(g_1, g_2) \gamma(g_1 g_2)
\eeq
for some factor $A(g_1, g_2)$. Of course, if $A(g_1,g_2)$ were
identically unity, then we are back to our familiar ordinary
representation (i.e., a linear homomorphism to a matrix group). It
turns out that $A(g_1, g_2)$ must obey a cocycle condition and is in
fact classified by the group cohomology $H^2(\Gamma, \IC^*)$. Our
orbifold group admits a projective representation iff
$H^2(\Gamma, \IC^*)$, dubbed {\bf discrete torision}, does not vanish.

In string theory, this is an old problem. It was realised in the first paper
on orbifolds by Dixon, Harvey, Vafa and Witten \cite{DHVW} that (in
the closed string sector) the partition function for the orbifold
theory can admit an ambiguity factor. In other words, in writing the
full partition function that includes the twisted sectors, one could
prepend the terms with a phase factor obeying certain cocycle
conditions as constrained by modular invariance. 
In the open string sector, this extra degree of freedom was realised
in \cite{opendis} to be precisely the possibility of discrete
torsion. 

Such liberty, wherever admissible,
gives us new classes of gauge theories that could differ markedly from
the zero discrete torsion case \cite{mydis}. Physically, what is
happening to the D-brane? There has been long investigated, notably by
Connes, Douglas, Schwarz, Seiberg and Witten, that there is an
underlying non-commutative structure in string theory \cite{noncom}. 
For the D-brane probe, if one turned on a background NSNS B-field
along the world volume, then the moduli space is actually expected to
be a non-commutative version of a Calabi-Yau space \cite{ncmoduli}.
This scenario is the physical realisation of discrete torsion. The
exponential of the B-field, as it was in the DHVW case as the
complexification of the K\"ahler form, corresponds to the phase
ambiguity.

A highly intuitive and visual way of studying gauge theories from
brane dynamics is the so-called {\bf Hanany-Witten setup} wherein
D-branes are stretched between configurations of
NS5-branes. Supersymmetry is broken according to the setup and the
world-volumes prescribe desired gauge theories \cite{HW}. The relative
motion of the branes provides the deformations and moduli in the
physics.

It is re-assuring that there is a complete equivalence between our
D-brane probe picture and the Hanany-Witten setups (and, actually,
also with geometrical engineering methods wherein D-branes wrap
vanishing cycles, a point to which we shall later return), cf.~e.g.,
\cite{Karch:equiv}. The mapping is through T-duality. The earliest
example was the realisation that T-duality of $\IC^2/\IZ_n$, the first
of the ADE quiver theories, gives $n$ NS5-branes placed in a ring; the
world-volume theory of D4-branes stretched between these branes, the
so-called elliptic model, is the $\cN=2$ A-type orbifold theory
discussed above. With the aid of orientifold planes, one could find
the brane setup of D-type orbifolds \cite{kapustin}. The three
exceptional cases, however, still elude current research.

In dimension three, the abelian orbifolds $\IZ_m \times \IZ_n$ can be
dualised to a cubic version of the elliptic model, appropriately
called brane boxes \cite{BBox}. Similar orientifold techniques have
been applied to other $\IC^3$-orbifolds \cite{ZD-ori}. The general
problem of constructing the Hanany-Witten
setup given an arbitrary orbifold group remains a tantalising issue
\cite{step}.

\setcounter{equation}{0}
\section{Gauge Theories, Moduli Spaces and \\
Symplectic Quotients}
Having expounded upon some details on orbifolds and seen intricate
mathematical structures that also manufacture various gauge theories
in four dimensions, we are naturally lead to wonder whether a general
approach is possible; i.e., given any singularity, how does one
reconstruct the gauge theory on the D3-brane world volume? We are in
desperate need of ``the method,'' and being of the Cartesian School, I 
quote,
``Car enfin La M\'ethode qui enseigne \`a
suivre le vrai ordre,
et \`a d\'enombrer exactement toutes les circonstances de ce
qu'on cherche, contient tout ce qui donne de la certitude aux
r\`egles d'arithm\'etique''
(R.~Descartes, {\it Discours Sur La M\'ethode}). We shall see later
that these rules of arithmetic, ingrained into the computations of
algebraic geometry, will constitute algorithms that will help answer
our question above.

The converse of our question, i.e., to obtain the singularity given
the gauge theory, is a relatively simple one. Indeed, the vacuum
parametre space of the scalar (bosonic) matter fields of the gauge
theory is the so-called {\bf moduli space} (we will give a more
precise definition later). As emphasised in the introduction, and we
re-iterate here, by our very construction, this vacuum moduli space,
because our D3-brane is a point to the transverse Calabi-Yau
threefold, is exactly the threefold. In other words, in local variables,
the moduli space $\cM$ of the world-volume gauge theory is the
affine \coords of the Calabi-Yau singularity $S$.

In the case of the abovementioned ADE $\cN=2$ theories, the moduli
space, by the Kronheimer-Nakajima construction \cite{KN}, is a
generalisation of the ADHM instanton moduli space. Their result, is a
hyper-K\"ahler quotient. In general, the moduli space can be
constructed as a so-called {\bf quiver variety}. We will see extensive
examples of this later.

The lesson I wish to convey is that there is a bijection between the
four-dimensional SUSY world-volume gauge theory and the Calabi-Yau
singularity. We shall adopt an algorithmic outlook.
To proceed from the physics to the mathematics is the
calculation outlined in the previous two paragraphs as we compute the
moduli space of the gauge theory; this we call the {\bf Forward
Algorithm}. To proceed from the mathematics to the physics is our
desired question as we extract the gauge theory given the
geometry of the Calabi-Yau singularity, this we call the {\bf Inverse
Algorithm}.

\subsection{Quiver Gauge Theory}\label{s:quivergauge}
For the mathematicians in the audience let me assume a moment of
attempted rigour and define what we have been meaning by our $\cN=1$
four-dimensional super-Yang-Mills gauge theory. For our purposes, {\it
a world-volume gauge theory is a (representation of a)
finite labelled graph (quiver) with relations.} It is finite because
there are a finite number of nodes and arrows, representing gauge
factors and matter fields. It has a label $\{n_i \in \IZ_+\}$ for the
nodes, signifying the dimensions of vector spaces $\{V_i\}$ each of
which is associated with a node. The gauge group is $\prod\limits_i
SU(n_i)$. The gauge fields are then self-adjoining arrows
$\Hom(V_i,V_i)$  while the matter fields are bi-fundamentals
fermions/bosons and are arrows $X_{ij} \in
\Hom(V_i,V_j)$ between nodes. 
In addition, the matter content must be {\bf anomaly free}.
This is a condition which ensures that the quantum field theory is
well-defined. For the quiver with adjacency matrix $a_{ij}$ and node
labels $n_i$, the condition reads
\beq\label{anom}
(a_{ij} - a_{ji}) n_i = 0.
\eeq
In other words, the ranks of the gauge groups must lie in the
nullspace of the antisymmetrised adjacency matrix.
The above data then specify the matter content.

Finally,
there are relations which arise from interaction terms in the field
theory. These are algebraic relations satisfied by the fields $X_{ij}$.
These relations arise from a (polynomial)
{\bf superpotential} $W(\{X_{ij} \})$, the
generalisation of a potential in ordinary field theory. 
Indeed, as it is in the case of classical mechanics, the vacuum is
prescribed by the minima of the (super)potential. 
In other words, the relations come
from the critical points
\beq\label{diffW}
\diff{W}{X_{ij}} = 0.
\eeq

Indeed, (the supersymmetric extension of)
our Standard Model is a generalisation of this structure above.
The Holy Grail of string theory is to be able to obtain the
Standard Model's (generalised) quiver from a unique compactification geometry.
As a hypothetical example, the quiver below is a $U(1)^2 \times SU(2)
\times SU(3)$ gauge theory with 8 matter fields
$X_{\alpha=1,\ldots,8}$. These fields carry
gauge indices: $(X_{\alpha})^i_j$ being a $SU(n_i) \times SU(n_j)$
bifundamental.
Relations could be such polynomial constraints as
$(X_1)^i_j (X_2)^j_k = (X_5)^i_m (X_4)^m_k$.
\beq\label{quivereg}
\ba{ccc}
\mbox{Adjacency Matrix} & \mbox{Incidence Matrix} & \\
{\small
\ba{c|cccc} & A & B & C & D \cr \hline A & 0 & 1 & 1 & 2 \cr
	B & 0 & 0 & 0 & 2 \cr C & 0 & 0 & 0 & 1 \cr
	D & 1 & 0 & 0 & 0
\ea
}
&
{\small
\ba{c|cccccccc} & 1 & 2 & 3 & 4 & 5 & 6 & 7 & 8 \cr \hline
A &-1 &-1 &-1 &0 &-1 &-1 &-1 &1 \cr
B &1 &0 &0 &0 &0 &0 &0 &0 \cr
C &0 &0 &0 &-1 &1 &0 &0 &0 \cr
D &0 &1 &1 &1 &0 &1 &1 &-1
\ea
}
&
\ba{c}\psfig{figure=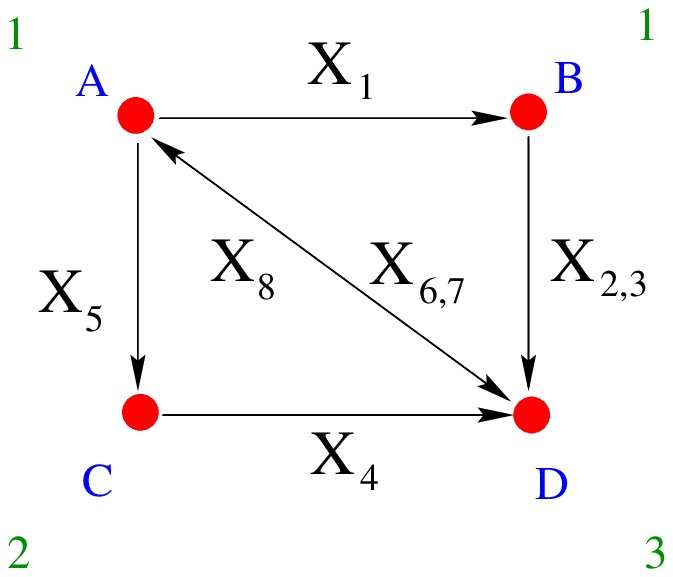,width=2in}\ea
\ea
\eeq
We have used two equivalent methods of encoding the quiver in
\eref{quivereg}, the adjacency matrix introduced in \sref{s:quiver}
and a rectangular {\bf incidence matrix} $I$ whose columns index the arrows
and the rows, the nodes, such that the
$\alpha$-th arrow from node $i$ to $j$
receives a $-1$ in position $I_{i\alpha}$ and a 1 in position
$I_{j\alpha}$ and zero elsewhere.

This more axiomatic approach above is not a self-indulgence into
abstraction but rather a facilitation for computation. 
In summary, our algorithmic perspective is as follows:
\[
\ba{|ccl|}
\hline
\mbox{PHYSICS: Gauge Data} & 
\ba{c}
\mbox{Forward Algorithm}\\
\rightleftharpoons\\
\mbox{Inverse Algorithm}
\ea
& 
\mbox{MATHEMATICS: Geometry Data} \\ \hline
\Updownarrow		&	&	\Updownarrow \\
\mbox{QUIVER}&\rightleftharpoons& \mbox{Intersection Theory, etc.} \\
\hline
\ea
\]

\subsection{An Illustrative Example: The Conifold}\label{s:conieg}
As a real example let us look at a famous case-study of a gauge theory
corresponding to a well-known singularity: the
so-called {\bf conifold} singularity; it is a Calabi-Yau threefold
singularity whose affine \coords are given by a hyper-surface in
$\IC^4$:
\beq\label{coni}
\{ u v - z w = 0 \} \subset \IC^4.
\eeq
The world-volume gauge theory of a stack of $N$ D3-brane probes was
shewn in \cite{Kle-Wit} to have 4 bi-fundamental fields $A_{1,2},
B_{1,2}$ with superpotential $W$ as follows:
\beq\label{conithy}
\ba{l}\psfig{figure=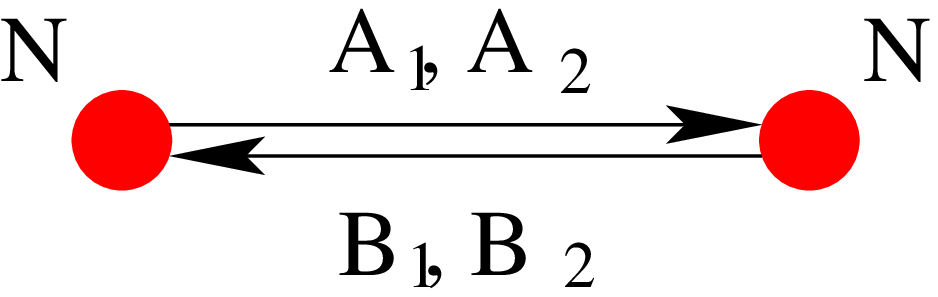,width=1.5in}\ea
\ba{ccc}
	& SU(N) & SU(N) \\
A_{i=1,2} & \fund & \antifund \\
B_{j=1,2} & \antifund & \fund
\ea
\qquad W = \Tr(\epsilon_{il} \epsilon_{jk} A_i B_j A_l B_k).
\eeq

For simplicity, take $N=1$, i.e., let all gauge factors be
$U(1)$. Then $W=0$ and no extra conditions \eref{diffW} are
imposed. The gauge invariant operators, i.e., combinations of fields
that carry no net gauge index are easily found: these are simply the
{\bf closed loops} in the quiver diagram. Here, they are 
$u = A_1 B_1,~v=A_1 B_2,~z=A_2 B_1,~w=A_2 B_2$. These scalars must
parametrise the vacuum moduli space. Since $W$ gives no further
relations here, we merely have a single relation amongst them, viz.,
$uv-zw=0$, precisely the affine equation \eref{coni}. This is what we
mean by having the gauge theory vacuum being the Calabi-Yau
singularity, the conifold. 
What we have just performed, was the Forward Algorithm.
In general, for $N>1$, one obtains an
$N$-th symmetrised product of the conifold.

%
\subsection{Toric Singularities}\label{s:toricsing}
With the above example let us launch into our next class of
singularities of \fref{f:state}, the toric cases. Whereas orbifolds
are the next best thing to flat space, toric varieties are the next
best thing to tori. Began in the 1970's, these spaces have been
extensively used in the early days of constructing Calabi-Yau
manifolds. Even completely outside the realm of string theory, many
gauge theories have their classical moduli spaces being toric
varieties, such as the conifold example in \sref{s:conieg}.
The Forward Algorithm and some of the Inverse Algorithm
for toric singularities, among a host of results on D-brane
resolutions, have been
beautifully developed in \cite{AReso,DGM,Horizon,chris,DD} 
based on the gauged linear sigma
model techniques of \cite{glsm}. The Inverse Algorithm in this context
was formalised in \cite{mytoric}.

Whereas we have shewn in \sref{s:orb} that the geometry of 
orbifolds is essentially captured by the representation theory of
the finite group, for the toric singularities, the geometry data will
be encoded in certain combinatorial data. I must point out that there
is a limitation to the gauge theory data due to the inherent Abelian
nature of toric varieties. The algorithms can only treat product $U(1)$
groups, i.e., $n_i = \dim V_i = 1$ for all the labels. Moreover, the
relations imposed on the arrows must be in the form $\prod_\alpha
X_\alpha = \prod_\beta X_\beta$, the so-called generators of {\bf
monomial ideals} \cite{sturm}. One could get higher rank gauge groups
by placing stacks of branes but the algorithms we present below will
capture only the Abelian information.

%
\subsubsection{A Lightning Review on Toric Varieties}\label{s:toricintro}
More to set nomenclature than to provide an introduction, let me
outline the rudiments of toric geometry; the audience is referred to
the excellent texts \cite{torictext}. An $r$ complex dimensional 
affine toric variety is
specified by a integer cone $\sigma$ in an integer lattice $\IZ^r$. To 
extract the variety from $\sigma$ we proceed as follows:
\begin{enumerate}
\item Find the dual cone $\sigma^\vee$, i.e., the set of vectors $w$ such
that $v \cdot w \ge 0$ for all $v \in \sigma$;
\item Find the intersection $\sigma^{\vee} \cap \IZ^r \leadsto$
	finitely generated semigroup $S_{\sigma}$;
\item Find the polynomial ring $\IC[S_{\sigma}]$ by exponentiating the
	\coords of $S_\sigma$;
\item The maximum spectrum (i.e., set of maximal ideals)
	of  $\IC[S_{\sigma}]\leadsto$ the toric variety.
\end{enumerate}
Compact toric varieties correspond to gluing these affine cones into a
fan, but we will, of course, be interested only in the local patches
and thus will focus only on cones. As a concrete example, consider
the following (I point out that instead of cones, physicists often use
the notation of simply drawing the lattice generators of the
cone. Therefore, in this notation, the toric diagram is simply a
configuration of lattice points marked below):
\beq\label{toriceg}\ba{ll}
\ba{c}\psfig{figure=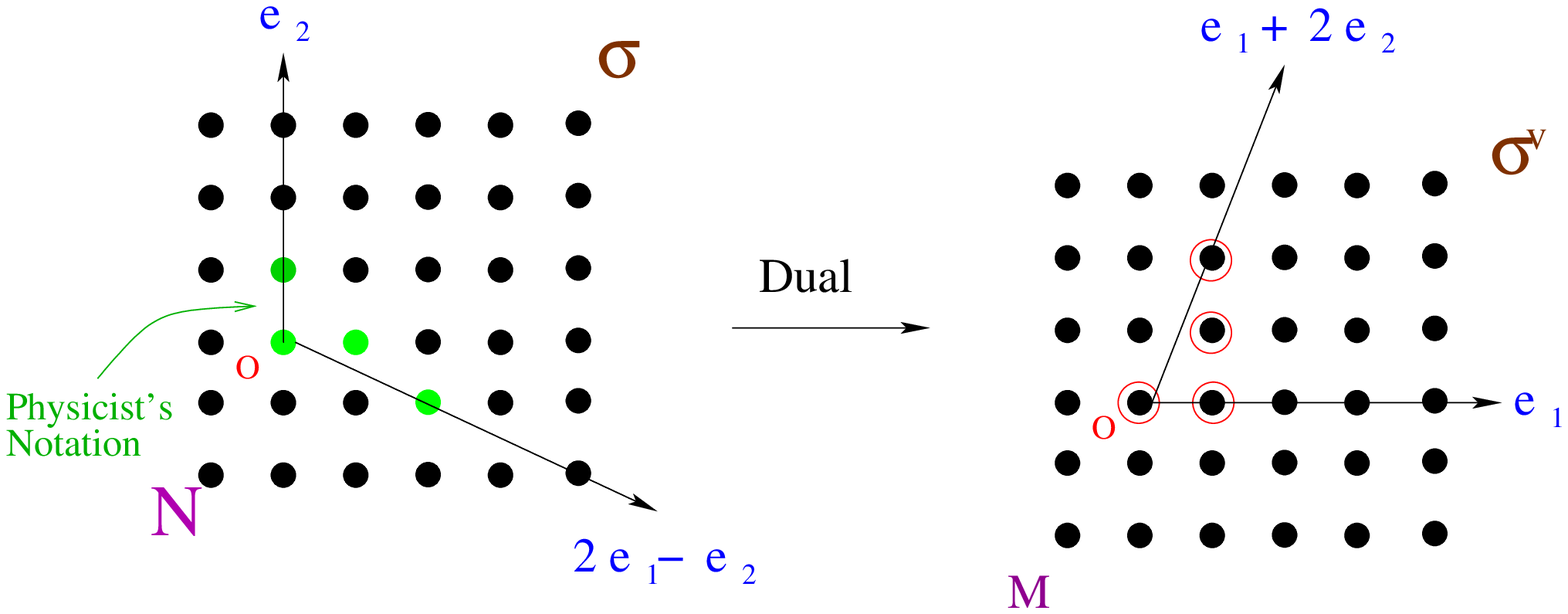,width=4.5in}\ea
\ba{c}
S_\sigma = \langle (1,0); (1,1); (1,2) \rangle \Rightarrow\\
\IC[S_\sigma] = \IC[X^1Y^0,X^1Y^1,X^1Y^2] \\
\qquad \equiv \IC[u,v,w] /(v^2-uw) \Rightarrow\\
{\rm Spec}_{Max}\left(\IC[S_\sigma]\right) = \IC^2/\IZ_2
\ea
\ea\eeq
The above is our familiar orbifold $\IC^2/\IZ_2$; I have explicitly shewn
the (2-dimensional) cone $\sigma$, its dual $\sigma^\vee$, the
semi-group $S_\sigma$, the polynomial ring $\IC[S_\sigma]$ as well as
how its maximal spectrum leads to the defining affine equation of the
orbifold. In fact, {\it all Abelian orbifolds are toric varieties}, 
a piece of information, shewn in \fref{f:state}, 
that will be of great use to us later.

%
%
\subsubsection{Witten's Gauged Linear Sigma Model (GLSM)}\label{s:glsm}
Witten in \cite{glsm} gave a physical perspective on toric
varieties. The prescription in the previous subsection
in fact has a field theory analogue. Even though it was originally used
in the context of two-dimensional sigma models, it gives us the right
approach to the Forward Algorithm. Here is an outline of Witten's method.
Take $x_i$ as \coords of $\IC^q$, and a $\IC^{*(q-d)}$-torus action
\beq
\lambda_a : x_i \rightarrow \lambda_a^{Q_i^a} x_i,
\eeq
where $\lambda_a \in \IC^*$ and 
$Q_{i=1,\ldots,q}^{a=1,\ldots,q-d}$ is a $q \times (q-d)$ integer
matrix. The (symplectic) quotient of $\IC^q$ by this action gives a
$d$-dimensional space. This is our desired toric variety. There could
be relations among the row vectors of the matrix $Q_i^a$, viz.,
\beq\label{Qmat}
\sum_i Q_i^a v_i = 0~\forall~a.
\eeq
In our notation in \sref{s:toricintro}, the vectors $v_i$ defines the
cone $\sigma$ while $Q_i^a$ is the semi-group $S_\sigma$ of the
lattice points in the dual cone. Witten's insight was to realise
$Q_i^a$ as charges of fields in a $U(1)^q$ QFT; the final affine
\coords of the variety, in the spirit of \cite{cox}, are homogeneous
\coords $\{z_a = \prod_i x_i^{v_i^a} \}$. This charge matrix, which
encodes all the information of the variety, will be
key to our algorithm.

Another crucial property of toric varieties that we need is the
so-called {\bf moment map}. A toric variety is naturally equipped with
a symplectic form, and with such, is always armed with such a map. I
will not bore the audience with the formal definition, which
essentially is a mapping that takes the variety to it associated toric
diagram (polytope). In Witten's language, this map is simply
\eref{Qmat}. What is convenient is that to perform a K\"ahler
resolution, i.e., $\IP^1$-blowup, of the singularity, one merely
changes the right-hand side from 0 to some parametres $\zeta_a \in
\IR_+$, known as {\bf Fayet-Iliopoulos} parametres. 
The charge matrix $Q_i^a$ together with these parametres
completely specify the resolved toric manifold. Graphically, this
desingularisation corresponds to node-by-node deletion from the toric
diagram, each deletion signifying a $\IP^1$-blowup \footnote{As
mentioned in the example in the previous subsection, the toric diagram
in physicists' notation is a configuration of points. In the cone
language of mathematicians, desingularisation corresponds to a process
called stellar division of the cone.}.

\subsection{The Forward Algorithm}
Endowed with the knowledge of some requisite rudiments on toric
varieties, we can now proceed to the Forward Algorithm. We must
re-cast the procedure of solving for the vacuum moduli space into the
above language of the gauged linear sigma model in the manner of
\cite{AReso,DGM,Horizon,chris,DD}. This not only greatly simplifies
and systematises our computation, but also will enable us to construct
gauge theories with the Inverse Algorithm.

The definition of gauge theory moduli space presented in
\sref{s:quivergauge}, for the case of toric singularities, 
can now be formalised to the following:
\begin{definition}\label{def:M}
The {\bf Moduli space} ${\cal M}$ of a $U(1)^k$ Quiver
Yang-Mills gauge theory with matter content given by incidence
matrix $d_{ia}$ and interactions given by monomial relations 
$\prod X_a^{i_a} = \prod X_b^{i_b}$
is the space of solutions to the following two
equations:
\ben
\item {\bf D-Term:} $D_i = \sum\limits_a d_{i a} |X_a|^2 = 0$;
\item {\bf F-Term:} $\prod X_a^{i_a} = \prod X_b^{i_b}$,
\een
where $i=1,\ldots k$, $a = 1, \ldots m$ with $k=$ \# Nodes and $m=$
\#arrows.
\end{definition}

Indeed, the fact that
all quiver labels are $1$ and that the F-terms generate
monomial ideals, as discussed earlier in \sref{s:toricsing}, 
is what we call the {\bf toric condition}. 
Comparing the definition of the D-term with \eref{Qmat}, we see that
it is precisely the moment map. We conclude therefore that the matter
content of the gauge theory specifies a toric variety. However, as we
learnt from the conifold example in \sref{s:conieg}, this is not
sufficient. One must also take the interactions, i.e., the
superpotential $W$, into account. The F-terms above are precisely the
critical points of $W$ shewn in \eref{diffW}. A primary task of
\cite{DGM} is to actually transform the F-terms into the form of
D-terms and to encode the gauge theory into a combined charge matrix
$Q_t$.

%
\subsubsection{Forward Algorithm for Abelian Orbifolds}\label{s:fa-orb}
Recalling that all Abelian orbifolds are toric varieties, what better
place to start indeed than such an example. In point form let us
outline the procedure, following the notation of \cite{DD}:
\begin{enumerate}
\item Take $\IC^3 / \Gamma$; the Abelian finite subgroups $SU(3)$
	are $\Gamma = \IZ_n$ or $\IZ_n \times \IZ_m$ of
	order $r$;
\item The matter content is simply the McKay quiver obtained from
	${\bf 3} \otimes R_i = \bigoplus\limits_i^r a_{ij} R_j$, with
	$\#$nodes $=r$ and $\#$arrows $ m = 3r$. From the adjacency
	matrix $a_{ij}$ one obtains the 
	incidence matrix $d_{r \times m}$ describing the charges for
	the D-terms; 
\item Solving the F-Terms gives us
	$X_a = \prod\limits_j^{r+2} v_j^{K_{aj}}$ where $K_{m \times
	(r+2)}$ is an integer matrix of exponents
	and generates a convex polyhedral cone $M
	\simeq \IZ^{r+2}$. It fits into the sequence
	\beq
	0 \rightarrow R \rightarrow \IZ^{3r}
	\stackrel{K^t}{\rightarrow} M \rightarrow 0,
	\eeq
	with a sequence for $N \simeq \IZ^c$, the dual cone
	to $M$ as
	\beq
	0 \rightarrow S \stackrel{Q^t}{\rightarrow} \IZ^{c}
	\stackrel{T}{\rightarrow} N \rightarrow 0;
	\eeq
\item Indeed, adding the D-Term constraints forms the moduli space 
	\[
	{\cal M} \simeq \{\mbox{F-Term solution}\}//U(1)^r.
	\]
	Actually one can remove one over-all $U(1)$ and consider the
	matrix $\Delta_{(r-1) \times m}$ which is $d_{r \times m}$
	with a row deleted;
\item Factor $\Delta : \IZ^m \rightarrow \IZ^{r-1}$ into $V \circ p$,
	with $V : M \rightarrow \IZ^{r-1}$ and $p : \IZ^m \rightarrow
	M$. Combining with above gives an complex of sequences
{\small
\diagram[Postscript=dvips]
&&& 0 &&&&&& \\
& & & \uTo_{} & & & & & & \\
& 0 & \rTo_{} & N & \rTo^{} & \left(\IZ^m\right)^* & \rTo^{} & 
R^* & \rTo^{} & 0 \ ; \\
&   &         & \uTo^T\dTo_{U^t} & \luTo^{V^t} & \uTo_{\Delta^t} & 
& & & \\
& & & {\IZ}^c &\lTo^{U^tV^t} & \left(\IZ^{r-1}\right)^* & & 
& & \\
& & & \uTo^{Q^t} & & & & & & \\
& & & S & & & & & & \\
& & & \uTo & & & & & & \\
& & & 0 & & & & & & \\
\enddiagram
}
\item The final toric data is given by the integer matrix $G_{3 \times c}$
such that
{\small
\diagram
0&\rTo&S\oplus\left(\IZ^{r-1}\right)^*&
\rTo^{Q_t := \mat{Q^t \cr (VU)^t}}&{\IZ}^c&
\rTo&G&\rTo 0 \ .
\enddiagram
}
The matrix $Q_t$ is the total charge matrix, combining the D-terms and
F-terms into a single moment map;
\item Associate each column of $G$ as a GLSM field and one sees that
there is {\bf repetition}, i.e., multiple GLSM fields are associated
to a single node in the toric diagram. 
\end{enumerate}

As a specific case, consider $\IC^3 / (\IZ_2 \times \IZ_2)$, which has 
gauge theory data
\beq
\ba{ll}
\ba{l}\epsfxsize=5cm\epsfysize=5cm\epsfbox{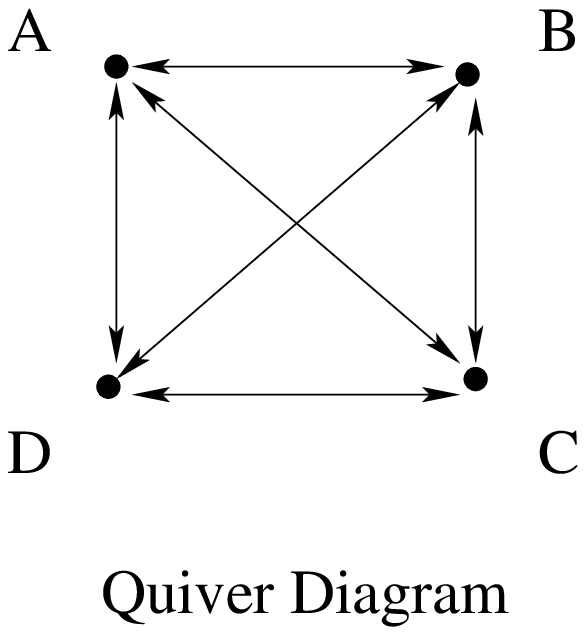}\ea
&
{\small
\ba{l}
[a_{ij}]_{4 \times 4} = \left( \matrix{ 0 & 1 & 1 & 1 \cr
                1 & 0 & 1 & 1 \cr 1 & 1 & 0 & 1 \cr 1 & 1 & 1 & 0
                \cr} \right)
\Rightarrow  [d_{ia}]_{4 \times 12}
\\
\begin{array}{ll}
W= &X_{AC}X_{CD}X_{DA} - X_{AC}X_{CB}X_{BA} + \\
	&X_{CA}X_{AB}X_{BC}- X_{CA}X_{AD}X_{DC} + \\
	&X_{BD}X_{DC}X_{CB} - X_{BD}X_{DA}X_{AB}- \\
	&X_{DB}X_{BC}X_{CD},
\end{array}	
\\
\frac{\partial W}{\partial X_a} = 0 \leadsto 12 \mbox{~F-terms}
\ea
}
\ea
\eeq
Applying the Forward Algorithm finally gives us the $G$ matrix
encoding the toric diagram, which we indeed recognise to be that of the
desired moduli space $\IC^3/(\IZ_2 \times \IZ_2)$, a generalisation of
our example in \eref{toriceg}. We have drawn the nodes according to the
columns of $G$ and have also marked the multiplicity when columns
repeat:
\beq\label{z2z2eg}
\ba{ll}
G = \left(
	\matrix{
	0 & 1 & 0 & 0 & -1& 0 & 1 & 1 & 1 \cr
	1 & 1 & 1 & 0 & 1 & 0 & -1& 0 & 0 \cr
	1 & 1 & 1 & 1 & 1 & 1 & 1 & 1 & 1}
\right) \Rightarrow
&
\ba{c}\epsfig{file=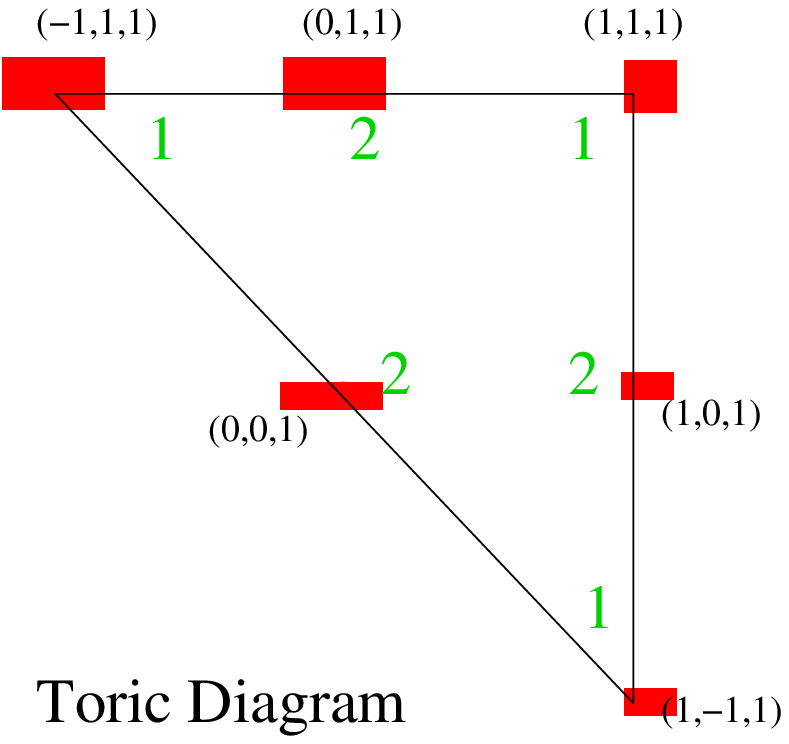,width=4cm}\ea
\ea
\eeq
One thing to note, of course, is that we have been able to draw
the toric diagram in
a plane even though we are dealing with a threefold. 
This is possible because we are dealing with Calabi-Yau singularities.
All the toric diagrams we
henceforth encounter will have this feature of co-planarity.

In summary then, we have a flow-chart that takes us from the physics
(quiver) data $(d,K)$ to the moduli space geometry (toric) data $G$:
\beq\label{flowchart}
\begin{array}{cccccc}
{\mbox{D-Terms}} 
\rightarrow d	& \rightarrow	&\Delta	& &&\\
	&	&\downarrow	&	&&\\

{\mbox{F-Terms}} 
\rightarrow K	& \stackrel{V \cdot K^t =
	\Delta}{\rightarrow}
		& V	 & &&\\
\downarrow	&	& \downarrow	& &&\\
T = {\rm Dual}(K)	& \stackrel{U \cdot T^t = {\rm
	Id}}{\rightarrow} & U & \rightarrow & VU&\\
\downarrow	&	&	&	& \downarrow\\
Q = [{\rm Ker}(T)]^t	&	& \longrightarrow & & Q_t =
	\left( \begin{array}{c}
		Q \\ 
		VU \end{array} \right)& \to {G =[{\rm Ker}(Q_t)]^t} \\
\end{array}
\eeq

%
%
\subsection{The Inverse Algorithm}\label{s:IA}
Our chief goal is to be able to obtain the gauge theory given the
geometry, i.e., to obtain the pair $(d,K)$ given $G$. Na\"{\i}vely,
I can simply trace back the arrows of flow-chart
\eref{flowchart}. However, the Forward Algorithm is a highly
non-unique and non-invertible process. Therefore, we must resort to a
{\it canonical method}. 

The method which we will use is the so-called
{\bf partial resolutions} used in \cite{chris,PRU} and formalised in
\cite{mytoric}. The procedure is as follows:
\begin{enumerate}
\item We first note that the toric diagram $D$ of any
Calabi-Yau threefold singularity embeds into the diagram $D'$ of the
Abelian orbifold $\IC^3/(\IZ_k \times \IZ_k)$ for large enough $k$. This
is because $D'$ is a triangle of lattice points (cf.~the example
\eref{z2z2eg} for the case of $k=2$). For example, the following is an
embedding of some given diagram $D$ into that of $k=3$:
\centerline{\psfig{figure=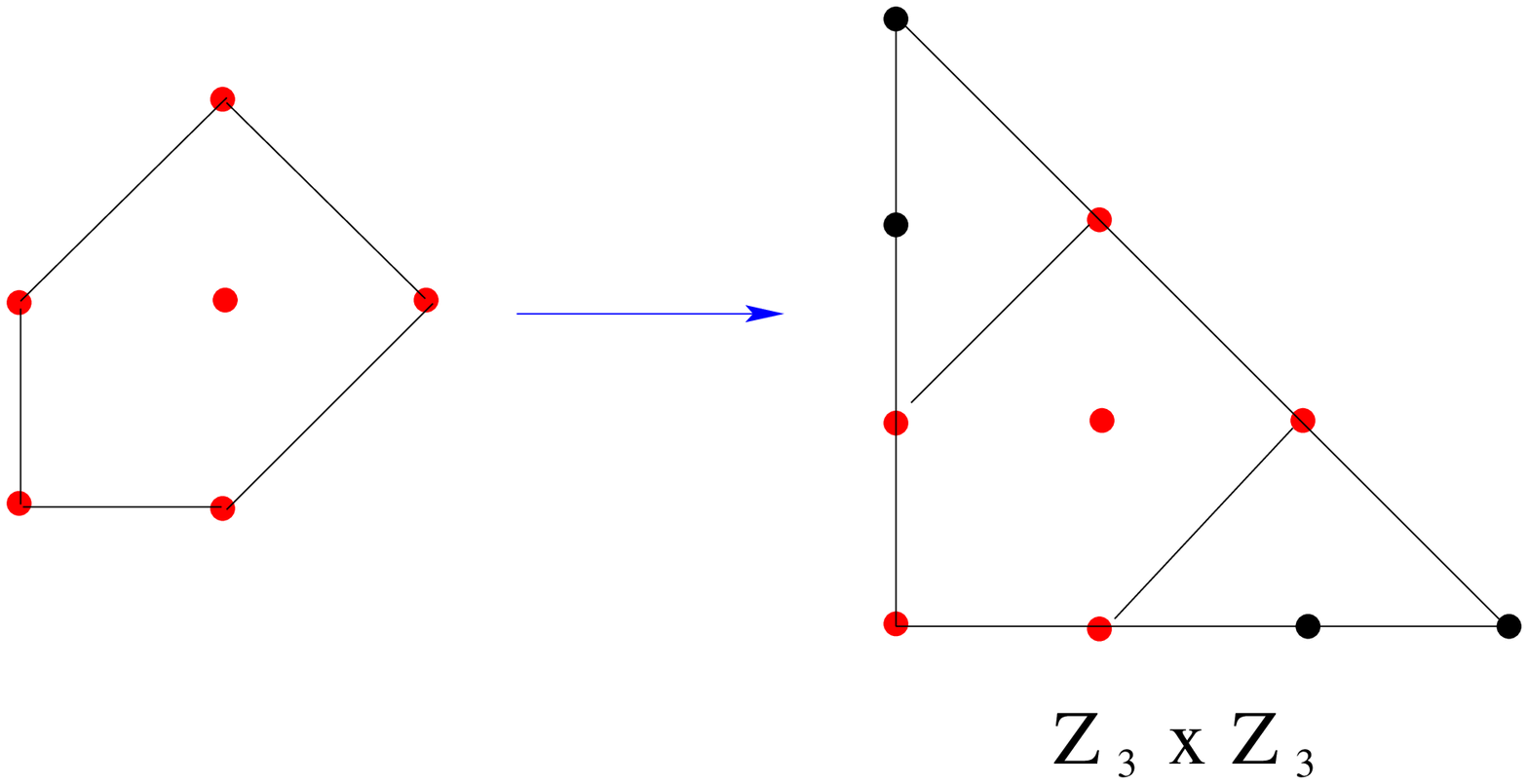,width=8cm}}

\item Now, $D'$ is an orbifold, thus the McKay analysis in \sref{s:orb}
conveniently gives us the gauge theory data. We can subsequently perform
the Forward Algorithm on $D'$ as was done in the example in
\sref{s:fa-orb}.

\item As mentioned at the end of \sref{s:glsm}, $D'$ can be
desingularised by node deletion. This stepwise blowup is called
partial resolution of the Abelian orbifold. In the GLSM picture, this
corresponds to the {\bf Higgsing}, i.e.,
acquisition of vacuum expectation values (VEV's)
of the GLSM fields. Therefore, one can partially resolve $D'$ 
until one reaches $D$. This is the crux of the algorithm, which
essentially is a game, since we are dealing with lattice polytopes, 
in {\it Linear Optimisation};

\item The gauge theory for $D$, our desired output, 
is then, by construction, a subsector of the theory for $D'$, via
stepwise acquisition of VEV's of GLSM fields.
\end{enumerate}

%
%
\subsection{Applications of Inverse Algorithm}
Supplanted with the method, one must apply to concrete examples. It is
expedient to introduce to the physicists in the audience a class of
singularities pertaining to {\bf del Pezzo surfaces}.
\subsubsection{del Pezzo Surfaces}
When His Grace Pasquale del Pezzo, Duca di Cajanello 
\cite{duca} took reposes from
political intrigues, his passions led to 
an important class of complex surfaces which now bears his name. Of
primary note is the fact that these surfaces constitute the two
complex dimensional analogues of the sphere $S^2$ in that they have
positive $c_1$. More strictly, they are the only surfaces with ample
anticanonical bundle, as $\IP^1 \simeq S^2$ is the only example in
dimension one. Therefore, they are the possible four-cycles which may
shrink in a Calabi-Yau threefold.

The first two members of the family are straight-forward: they are
$\IP^2$ and $\IP^1 \times \IP^1$. The remaining members are these two
surfaces 
with $k$ generic points thereupon blown up with $\IP^1$'s, with $k$ up
to 8. The case of $k=9$ is usually called the ninth del Pezzo surface
even though strictly it should not be so-named since its first Chern
class squares to 0. For this reason it is also called $\frac12 K3$.
The blowup relations within the family are as follows:
\beq\ba{ccc}
&(F_0 = \IP^1 \times \IP^1) &\\
&\downarrow&\\
(dP_0 = \IP^2) \rightarrow &dP_1 &\rightarrow dP_2 \rightarrow \ldots
\rightarrow dP_8 \rightarrow (dP_9 = \frac12 K3).
\ea\eeq

Of curious and McKay-esque interest is the second cohomology
$H^2(dP_k, \IZ)$ (q.v.~\cite{INV}). 
In $k=0$, $H^2(dP_0, \IZ)$ clearly consists only of
the single element $\ell$, which is the hyperplane class of $\IP^1
\subset \IP^2$. For $k>0$, each time there is a blowup, we introduce
an exceptional $\IP^1$-class $E_k$. Therefore
\beq\label{interdP}
H^2(dP_k, \IZ) = {\rm span}(\ell, E_{i=1,\ldots,k}), \qquad
E_i \cdot E_j = -\delta_{ij},~E_i \cdot \ell = 1,~\ell^2=0 .
\eeq
The first Chern class is given by
\beq
c_1(TdP_k) = 3 \ell - \sum_{i=1}^k E_i.
\eeq
We see indeed that $c_1(TdP_9)^2 = 0$ and $k=8$ is the last case for
which $c_1(TdP_k)^2 = 9-k > 0$. Remarkably, $H^2(dP_k, \IZ)$ in
\eref{interdP} is the root lattice of the exceptional Lie algebra 
$\IE_k$.

Finally, we note that $F_0, dP_{0,1,2,3}$ actually admit a toric
description. We are, of course, concerned not with the surfaces
themselves but affine cones over them which are Calabi-Yau
threefold singularities. Slightly abusing notation, we will also refer
to the affine cone by the same name as the surface. We now
apply the Inverse Algorithm in \sref{s:IA} to the affine
three-dimensional varieties $F_0, dP_{0,1,2,3}$.

%
\subsubsection{Application to Toric del Pezzo's}
Observing the toric diagrams of $F_0, dP_{0,1,2,3}$, we see that they
can all be embedded into $\IC^3 / (\IZ_3 \times \IZ_3)$:
\beq\label{dPtoric}
\centerline{\psfig{figure=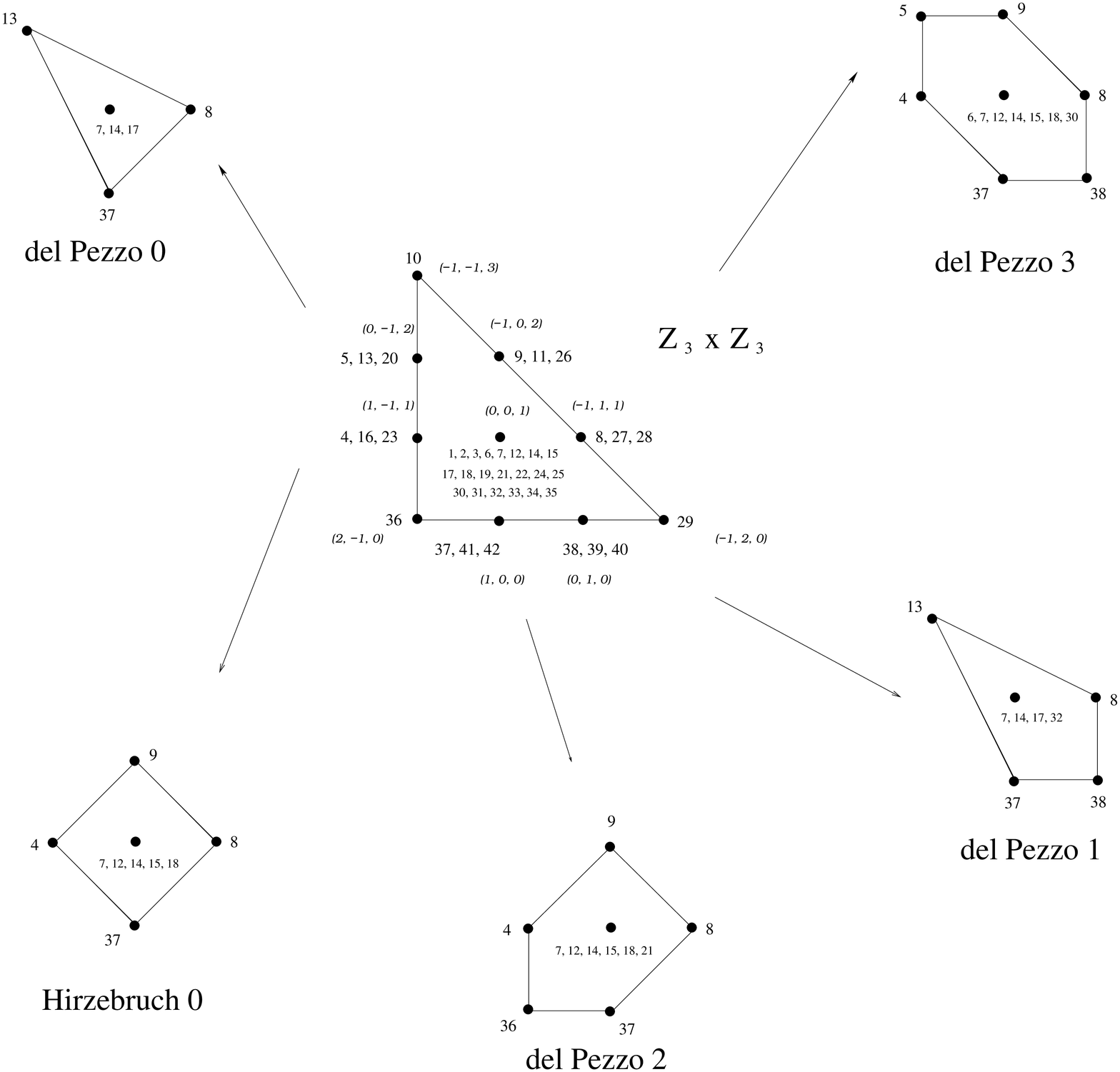,width=5in}}
\eeq
Running the Inverse Algorithm gives us the following quivers,
\beq\label{dPquiver}
\centerline{\psfig{figure=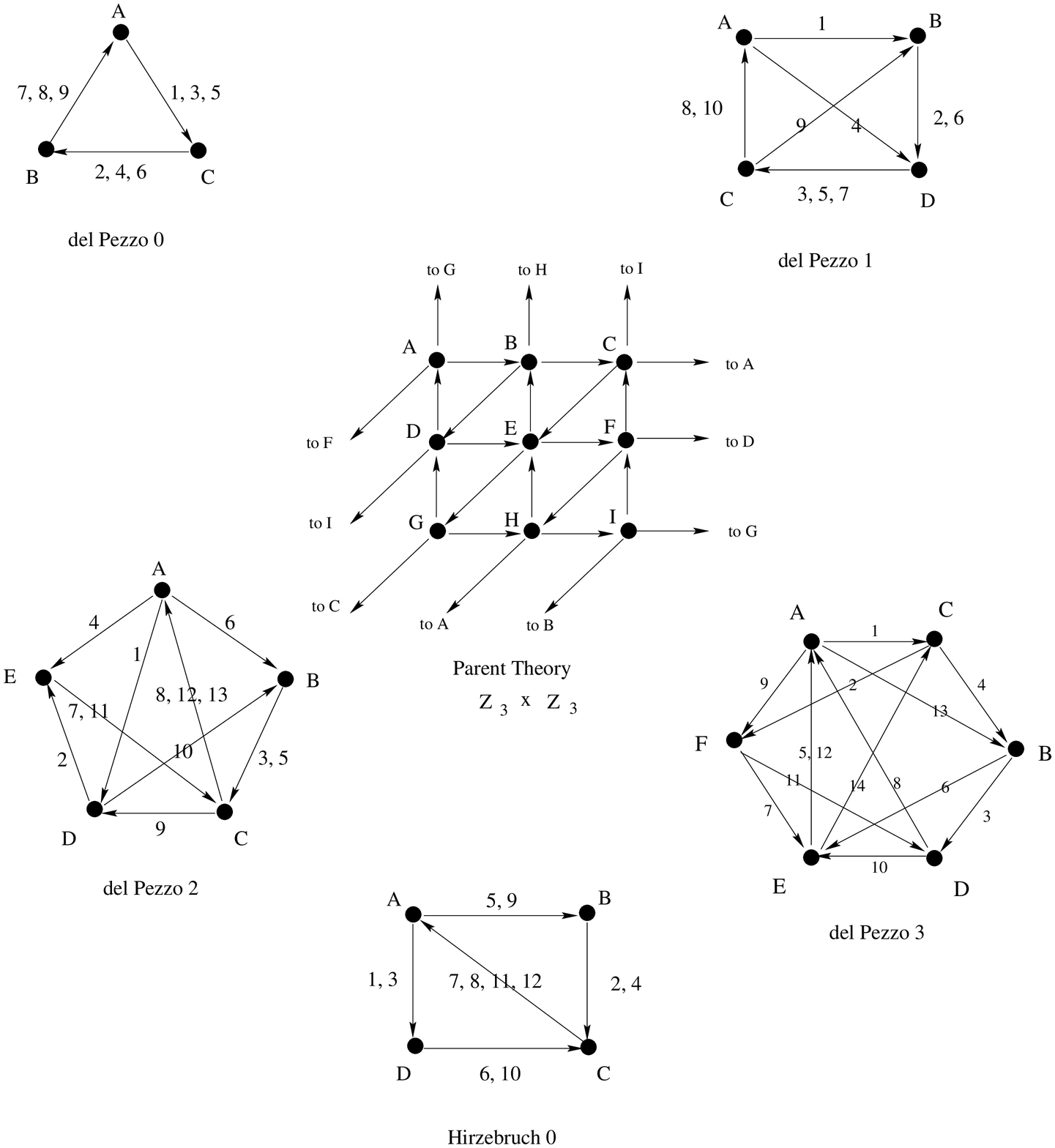,width=5in}}
\eeq
together 
with associated superpotentials which I will not present here.
As an immediate check, we know that the cone over $dP_0 = \IP^2$ is
$\IC_3 / \IZ_3$ as we could have seen from the toric diagram in
\eref{dPtoric}. 
In fact the crepant resolution of $\IC_3 / \IZ_3$ is the bundle
$\cO_{\IP^2}(-3)$. Hence, the quiver should simply be the McKay quiver
for $\IZ_3 \subset SU(3)$. Indeed the top-left quiver in
\eref{dPquiver} is consistent with this fact.

The Inverse Algorithm is general and we can apply it to any other toric
singularity. The only draw-back is that finding dual cones, a key to
the algorithm, is computationally intensive. I have been using the
algorithm in \cite{torictext} which unfortunately has exponential
running time. Of course, more efficient methods do exist but have not
yet been implemented in this context.
%
%
\subsection{Toric Duality, or, Non-Uniqueness: Virtue or Vice?}
Now, in \sref{s:IA} I emphasised that the Forward Algorithm in
\eref{flowchart} is highly non-invertible. This was why we resorted to
the canonical method of partial resolutions. However, can this vice be
turned into a virtue and a heaven be made from a Miltonian hell? If
the process is so non-unique, could we not manufacture a multitude of
$\cN=1$ world-volume gauge theories having the same moduli space? Can
one indeed have many pairs $(d,K)$ having the same $G$?
Of course, one needs to be careful. Not all theories having the same
$G$ are necessarily physical in that they can be realised as brane
probes. The Inverse Algorithm guarantees that the output is a physical
theory because it is a subsector of the well-known orbifold theory.

Therefore, one technique does ensure physicality. We recall from
\eref{s:fa-orb} that the final output of the Forward Algorithm, the
toric data $G$ has multiplicities and each node could correspond to
many different GLSM fields. The choice of which GLSM to acquire VEV,
so long as during the Inverse Algorithm the right nodes are deleted at
the end, is completely arbitrary. This ambiguity will constitute a
systematic procedure in finding physical theories having the same
toric moduli space. Other methods may be possible as well and the
audience is encouraged to experiment.

As a momentary diversion, let me intrigue you with an {\it observatio
curiosa}: if one performed the Forward Algorithm to various Abelian
orbifolds $\IC^3 / (\IZ_k \times \IZ_m)$, 
the multiplicities in $G$, which I will label in the
diagram below, actually exhibit the Pascal's Triangle (which my
ancestors, and I thank a member of the audience to have reminded me,
have dubbed the Yang-Hui Triangle, pre-dating Monsieur Blaise by about
500 years; however I shall refrain from evoking my namesake too much):
\beq
\centerline{\psfig{figure=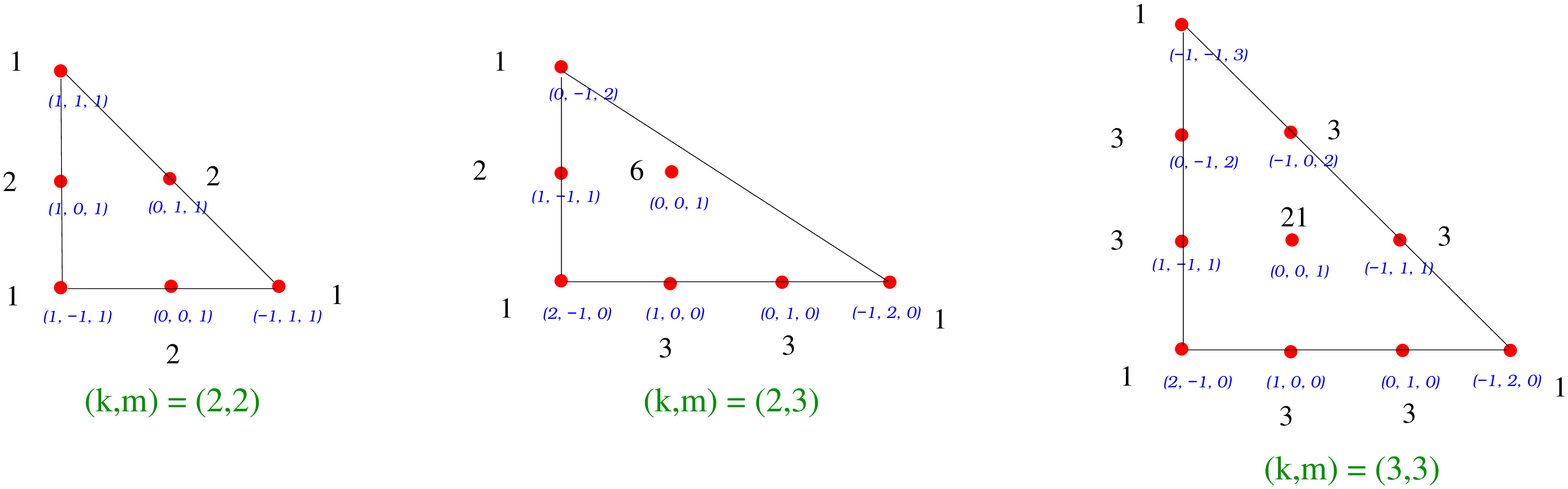,width=5in}}
\eeq
I have no explanation for this fact: the details of the algorithm are
not analytic and I know of no analytic determination of say,
generators of dual cones; thus I do not know how one might proceed to
prove this observation, which, curiously enough, is key to constructing
gauge theories with the same moduli space.

Bearing this insight in mind, we can re-apply the Inverse Algorithm
to \eref{dPtoric}, varying which GLSM fields to Higgs. The result is
the following, with 2 phases for $F_0$, 2 for $dP_2$ and 4 for $dP_3$;
Model I in each case refers to the theory in \eref{dPquiver}, though
drawn with more explicit symmetry (Model I of $dP_3$ for example, has
been re-arranged, by the good Hanany, in a flash of precipience, as if
the figure awakened a symbolism distilled into his very blood. ``The
Star of David, you see.'' he said calmly. 
``Ah! And Solomon's wisdom has inspired you
to get back to your roots.'' I jocundly replied. With a mischievous
glare in his eyes he shrugged his habitual non-chalance, motioned
to me with the palm of his hand, and gave his canonical response: 
``you said it.'')
\beq
\centerline{\psfig{figure=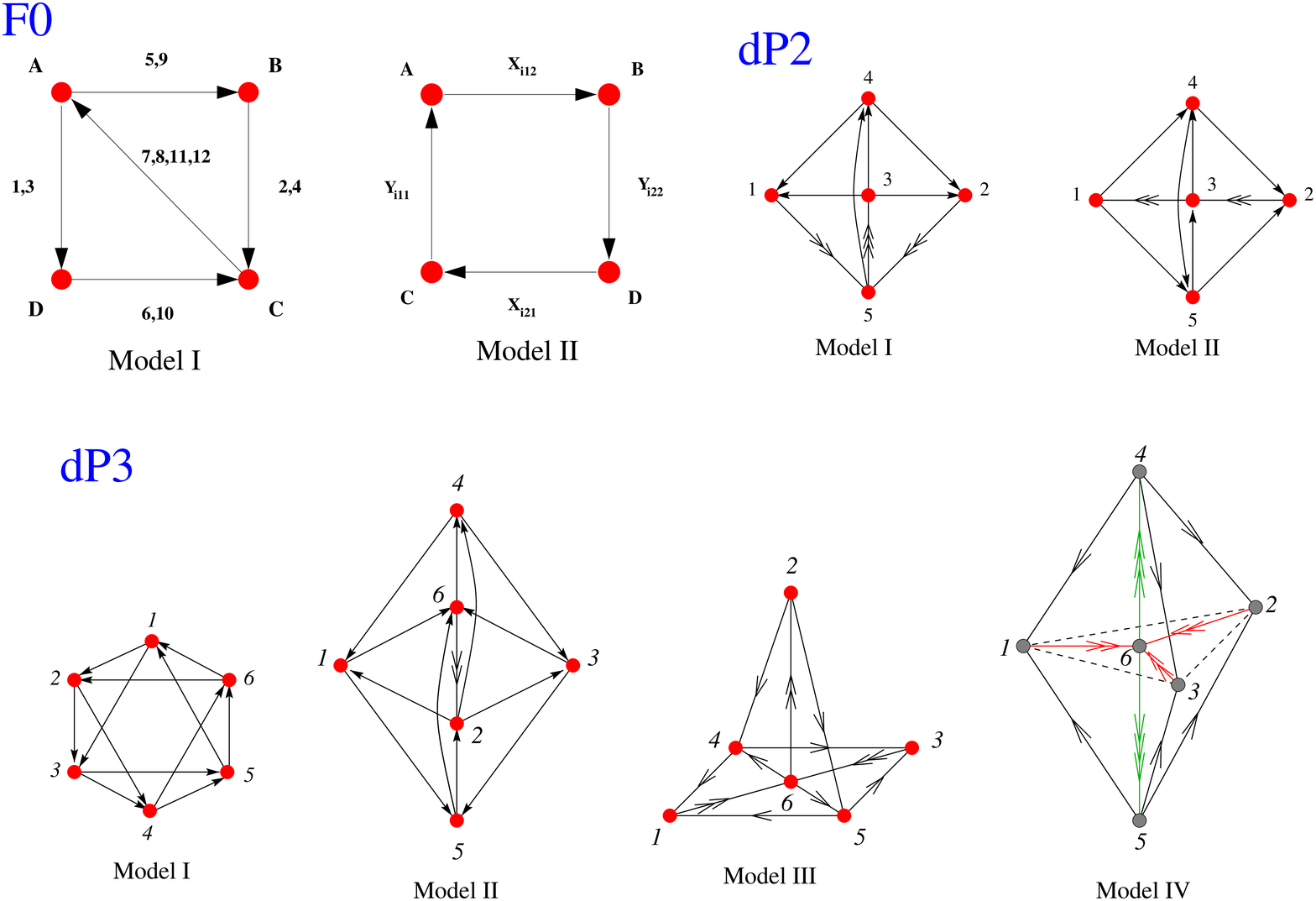,width=6in}}
\eeq

We will call this phenomenon of having different gauge theories having
the same IR moduli space {\bf toric duality}.

\subsubsection{What is Toric Duality?}\label{s:td=sd}
For the mathematicians let me briefly point out that in physics, a
duality between two QFT's is more than an identification of the
infra-red behaviour of the moduli space. It should be a complete mapping,
between the dual pair, of quantities such as the Lagrangian or
observables. Usually, the high-energy regime of one is mapped to the
low-energy regime of the other. Famous duality transformations exist
for QFT's with various supersymmetries. For $\cN=4$, there is the
Montonen-Olive duality, for $\cN=2$, Seiberg-Witten theory and for
$\cN=1$, there is Seiberg's duality.

Seiberg Duality is, in its original manifestation \cite{SD}, 
a duality (when $3N_c/2\leq N_f\leq 3N_c$) 
between the following pair of $\cN=1$ theories with a 
global chiral symmetry $SU(N_f)_L\times SU(N_f)_R$:
\beq
\hspace{-1cm}
\begin{tabular}{|c|c|c|}
\hline
	& Electric Theory	&	Magnetic Theory \\ \hline
Gauge Group &	$SU(N_c)$	&	$SU(N_f-N_c)$ \\ \hline
Fund.~Flavours	& $N_f$		&	$N_f$ \\ \hline
Matter Content  & 
$\ba{c|ccc}
        & SU(N_c)  &  SU(N_f)_L  &  SU(N_f)_R \\ \hline
Q       & \fund  &   \fund   &      1 \\
Q'      & \antifund &    1     &     \antifund \\
\ea$
&
$\ba{c|ccc}
        &SU(N_f-N_c)    &SU(N_f)_L      &SU(N_f)_R \\ \hline
q       & \fund         &\antifund      &1 \\
q'      & \antifund     &1              &\fund \\
M       &1              &\fund          &\antifund \\
\ea$ \\ \hline
Superpotential	& $W=0$	& $W = M q q'$ \\ \hline
\end{tabular}
\eeq

Such a dualisation procedure can be easily extended, node by node, to
our $\cN=1$ quiver theories. One might naturally ask whether toric
duality might be Seiberg duality in disguise. If so, our games on
integer programming would actually possess deep physical significance.
And so it was checked in \cite{td=sd} and also independently in the
lovely papers \cite{BP,Cachazo} that this is the case. 
We conjecture that {\em Toric Duality = Seiberg Duality 
for ${\cal N}=1$ SUSY theories with toric moduli spaces}. Of course,
this has no {\it a priori} reason to be so and indeed if we find toric
dual pairs that are not Seiberg dual then we would be led to the
interesting question as to what exactly this new duality would mean.

Now, ignoring the superpotential for a while and consider Seiberg
duality acting on the matter content.
Then, as an action on the quiver, it is the following set of rules:
\begin{enumerate}
\item   Pick dualisation node $i_0$. Define:
        $I_{in} :=$ nodes having arrows going into $i_0$; 
        $I_{out} := $ coming from $i_0$ and
        $I_{no} :=$ unconnected with $i_0$;
\item   rank($i_0$): $N_c \rightarrow N_f-N_c$ 
	($N_f=\sum\limits_{i\in I_{in}} a_{i,i_0}=
	\sum\limits_{i\in I_{out}} a_{i_0,i}$); 
\item $a^{dual}_{ij} = a_{ji}$
	$\qquad \mbox{ if either }i,j = i_0$;
\item $a^{dual}_{AB} = a_{AB} - a_{i_0 A} a_{B i_0}$
 	$\qquad \mbox{ for } A \in I_{out},~~B \in I_{in}$
\end{enumerate}

%
\subsubsection{Other Geometrical Perspectives}
It is now perhaps expedient to take some alternative geometrical
perspectives on our story. We shall see that the quiver transformation
rules at the end of \sref{s:td=sd} naturally arise as certain
geometrical actions.

\paragraph{The Mirror}
I can hardly give a talk on Calabi-Yau spaces without at least a
mention of mirror symmetry, a beautiful subject into which I shall not
delve here. For now, let us merely use it as a powerful tool.
Following the prescription of \cite{SYZ}, that mirror symmetry is
thrice T-duality, our configuration of D3-branes transverse to
Calabi-Yau threefold singularity is mapped to type IIA D6-branes
wrapping vanishing 3-cycles in the mirror Calabi-Yau threefold. We
have gained one dimension, both for the D-brane and for the homology,
each time we perform a T-duality: D3-branes thus become D6-branes and
the 0-cycles (recall that for the D3-probe the transverse Calabi-Yau
space is a point) become (Special Lagrangian) 3-cycles.

The mirror manifold of our affine singularities 
can be determined using the methods of
\cite{mirror}. This is known as ``local mirror symmetry.''
For example, the mirror of cones over del Pezzo surfaces is an
elliptic fibration over the complex plane.

In this context \cite{CV,soliton1,soliton2}, 
the quiver adjacency matrix is simply given by (up to
anti-symmetrisation and convention) the intersection of the vanishing
3-cycles $\Delta_i$:
\beq\label{aij-del}
a_{ij} = \Delta_i \circ \Delta_j.
\eeq
A convenient method of computing this intersection is to use the
language of {\bf $(p,q)$-webs} \cite{pq}. It turns out that each
3-cycle $\Delta_i$ wraps a $(p_i,q_i)$ 1-cycle in the
elliptic fiber in the mirror. The intersection number then simply
reduces to
\beq
a_{ij} = \Delta_i \circ \Delta_j = \det
\mat{p_i & q_i \cr p_j & q_j}.
\eeq
The configurations of $(p,q)$-webs can be a formidable simplification
in constructing the matter content for toric singularities because the
web itself is a straight-forward dual graph of the toric diagram. The
audience is referred to \cite{soliton1,soliton2,pq-stuff} 
which produce toric quivers very
efficiently. The only short-coming is that obtaining the
superpotential thus far escapes this simple approach.

Within this context of vanishing cycles, the rules for quiver Seiberg
duality at the end of \sref{s:td=sd} may look rather familiar. There
is an action in singularity theory \cite{arnold} known as
{\bf Picard-Lefschetz transformation} as we move a cycle $\Delta_i$
monodromically around a vanishing cycle $\Delta_{i_0}$. The cycle is
transformed as
\beq\label{PL}
\Delta_i \rightarrow \Delta_i - (\Delta_i \circ \Delta_{i_0})
	\Delta_{i_0}.
\eeq
Combining this with \eref{aij-del}, it was shewn in
\cite{soliton2,ito-k} 
that Seiberg duality is precisely a Picard-Lefschetz
transformation in the vanishing cycles of the mirror.

%
\paragraph{Helices and Mutations}
Moving back to the original Calabi-Yau singularity, there is yet
another description that has been widely used in the literature and
has been key in a general methodology. This is to regard the D-brane
probe as helices of {\bf coherent sheafs}
\cite{Cachazo,CV,soliton2,helix,Martijn,Herzog}. Once the helix is
constructed on the base over which our singularity is a cone, the
appeal of this approach is that it is completely general, being
unrestricted to orbifolds and toric singularities. One can obtain the
gauge data, both the matter and superpotential with the prescribed method
which we outline below. The caveat is that constructing the helix is a
rather difficult task for general singularities. However, for del
Pezzo surfaces, helices have known in the mathematics community
\cite{exceptional}. In fact, because of this fact, cones over the higher,
non-toric del Pezzo surfaces constitute the only known examples\footnote{
One could also use certain Unhiggsing techniques from pure field
theory \cite{BP, unhiggs}.
} of
non-orbifold, non-toric singularities for which probe gauge theories
have been constructed \cite{soliton1,Martijn}.

Let us begin with preliminaries. A collection of coherent
sheafs $F_i$, with a specified Mukai vector 
ch$(F_i) := ({\rm rk}(F_i), c_1(F_i), c_2(F_i))$, has a natural
intersection pairing given in terms of the Euler character:
\beq\label{aij-helix}
a_{ij}=\chi(F_{i},F_{j}) := 
\sum\limits_m (-1)^m \dim_{\IC} {\rm Ext}^m(F_i,F_j).
\eeq
In the case that the collection is {\bf exceptional}, i.e.,
$\ext^{m}(F_i,F_i) = \IC \delta_{m0}$ for all $i$, and
Ext$^m(F_i,F_j) = 0 $ for all $m$ if $j > i$ and 
for all but at most one $m$ if $i<j$, this intersection pairing
\eref{aij-helix} is precisely the adjacency matrix of the quiver
(cf.~\cite{mirror,Herzog}). The F-terms \cite{Cachazo} can be obtained
by successive Yoneda compositions, along closed cycles in the quiver,
of the Ext groups.

In this language, there is an action, perhaps inspired by biology, on
the exceptional collection, known as {\bf mutation}. Left ($L$) 
and right mutations ($R$)
with respect to the $l$-th sheaf in the collection proceed
as
\beq
\ba{ccc}
\ba{ccc}
\{F_{l},F_{l+1}\}&\mapsto& \{R_{l}F_{l+1},F_{l}\} \ ,
\\ &\mapsto&  \{F_{l+1}, L_{l+1}F_{l}\}\\
\ea
&
\mbox{s.t.}
&
\ba{ccc}
\mbox{ch}(R_{l}F_{l+1})&=&\mbox{ch}(F_{l+1})-\chi(F_{l},F_{l+1})
\mbox{ch}(F_{l})\,,\\
\mbox{ch}(L_{l+1}F_{l})&=&\mbox{ch}(F_{l})-\chi(F_{l},F_{l+1})
\mbox{ch}(F_{l+1}).
\ea
\ea
\eeq
This should once again be reminiscent of \eref{PL} and one could refer
to \cite{Herzog} to see how Seiberg
duality in this language is a mutation of the exceptional collection.
Of course, to fully explore the sheaf structure of the D-branes, as
we have secretly done above in the mutations, one must venture into
the derived category of coherent sheafs
(q.v.~e.g.~\cite{derived,Aspinwall} for marvelous  reviews). 
Computations in the context of del Pezzo singularities have been
performed in \cite{aspinmel}. 
Indeed, Seiberg duality in this context become
certain {\bf tilting functors} in $D^b(coh(X))$ \cite{tilt}.

With this digression I hope the audience can appreciate the rich and
intricate plot of our tale. From various perspectives, each with
her own virtues and harmartia, one can begin to glimpse the vista of
probe gauge theories. I have throughout these lectures emphasised upon
an algorithmic standpoint because of its conceptual simplicity and
transcendence above the difficulties and unknowns of geometry.

%
\setcounter{equation}{0}
\section{A Trio of Dualities: Trees, Flowers and Walls}
In this parting section, let me be brief, not so much that 
brevity is the soul of wit,
but that a bird's-eye-view over the landscape will serve more to
inspire than a meticulous combing of the nooks and crannies.
We will see how our excursion into gauge theories, algebraic geometry
and combinatorics will take us further into the realm of number theory
and chaotic dynamics.

%
\subsection{Trees and Flowers}
In \sref{s:td=sd} we have seen that a very interesting action can be
performed on our quivers, viz., Seiberg/toric duality. In general, the
rules therein can be applied to any quiver which may be
produced by the Inverse Algorithm. Indeed, though the said algorithm
will always produce quiver theories satisfying the toric conditions in
Definition \ref{def:M}, the Seiberg duality rules can well take us out
of these constraints. Of course, by definition of duality, these rules
must generate theories with the same moduli space which here happens to
be toric.

That said, one could dualise a given quiver {\it ad infinitum} with
respect to various choices of nodes at each stage. 
The result, is a dendritic
structure which we call the {\bf duality tree}. 
The node-labels, i.e.,
the ranks of the gauge group factors, amusingly enough, will always be
constrained to obey some classifying {\it Diophantine equation}.
Let us take the example of $dP_0$, the cone over the zeroth del Pezzo
surface $\IP^2$. 
The resolution space, as mentioned earlier, is simply ${\cal
O}_{\IP^2}(-3) \rightarrow \IC^3/\IZ_3$. The quiver was given in
\eref{dPquiver} and is a $U(1)^3$ theory in the toric phase. In the
diagram below, we leave the labels as well as the number of arrows
arbitrary, but satisfying the anomaly cancellation \eref{anom}.
The question is, what are the values of $n_{i=1,2,3}$ such that the quiver
may be obtained from a sequence of Seiberg dualities from the canonical
$\vec{n} = (1,1,1)$ quiver in the toric phase?
It turns out that \cite{CV,soliton2} they must satisfy the Diophantine
equation
\beq
n_1^2 + n_2^2 + n_3^2 = 3 n_1 n_2 n_3,
\eeq
which is known as the {\bf Markov equation}. We have shewn, in the
diagram below, the first 9
solutions of this equation in terms the quivers that can be obtained
from dualisation. The figure to the right is the duality tree
associated with $dP_0$. The trifurcating structure is due to the
$\IZ_3$ symmetry: at each stage of have a choice of 3 nodes to dualise.
\beq
\ba{cc}
\ba{l}
\epsfxsize=3cm\epsfbox{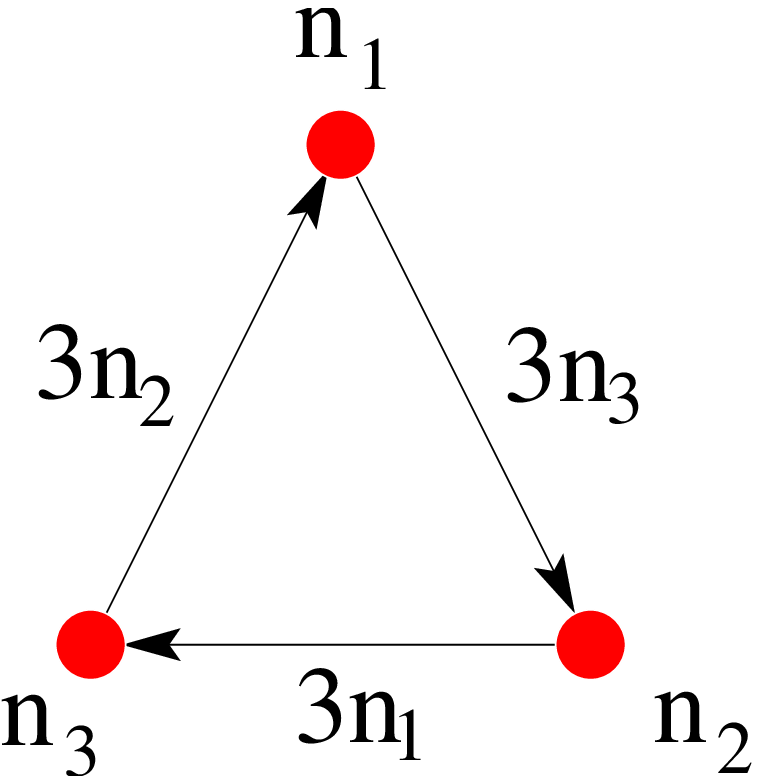} \\
\mbox{Markov Eq.~~~}n_1^2 + n_2^2 + n_3^2 = 3 n_1 n_2 n_3 \\
\epsfxsize=10cm\epsfbox{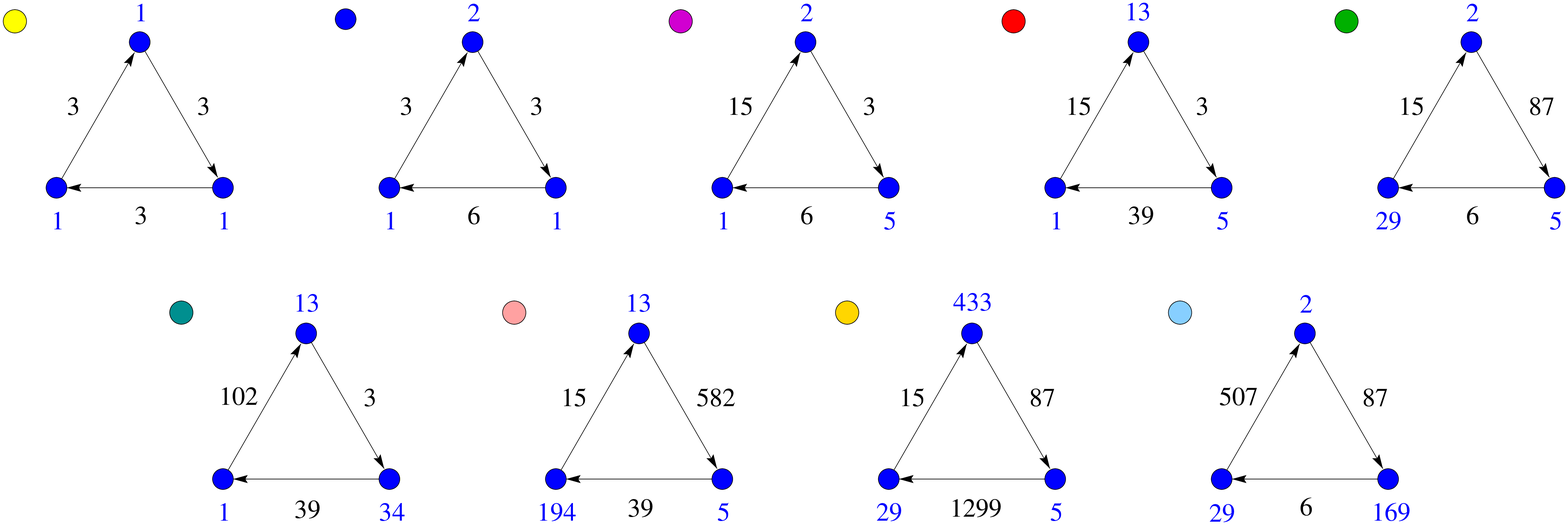}\\
\ea
&
\hspace{-1cm}
\ba{c}
\epsfxsize=7cm\epsfbox{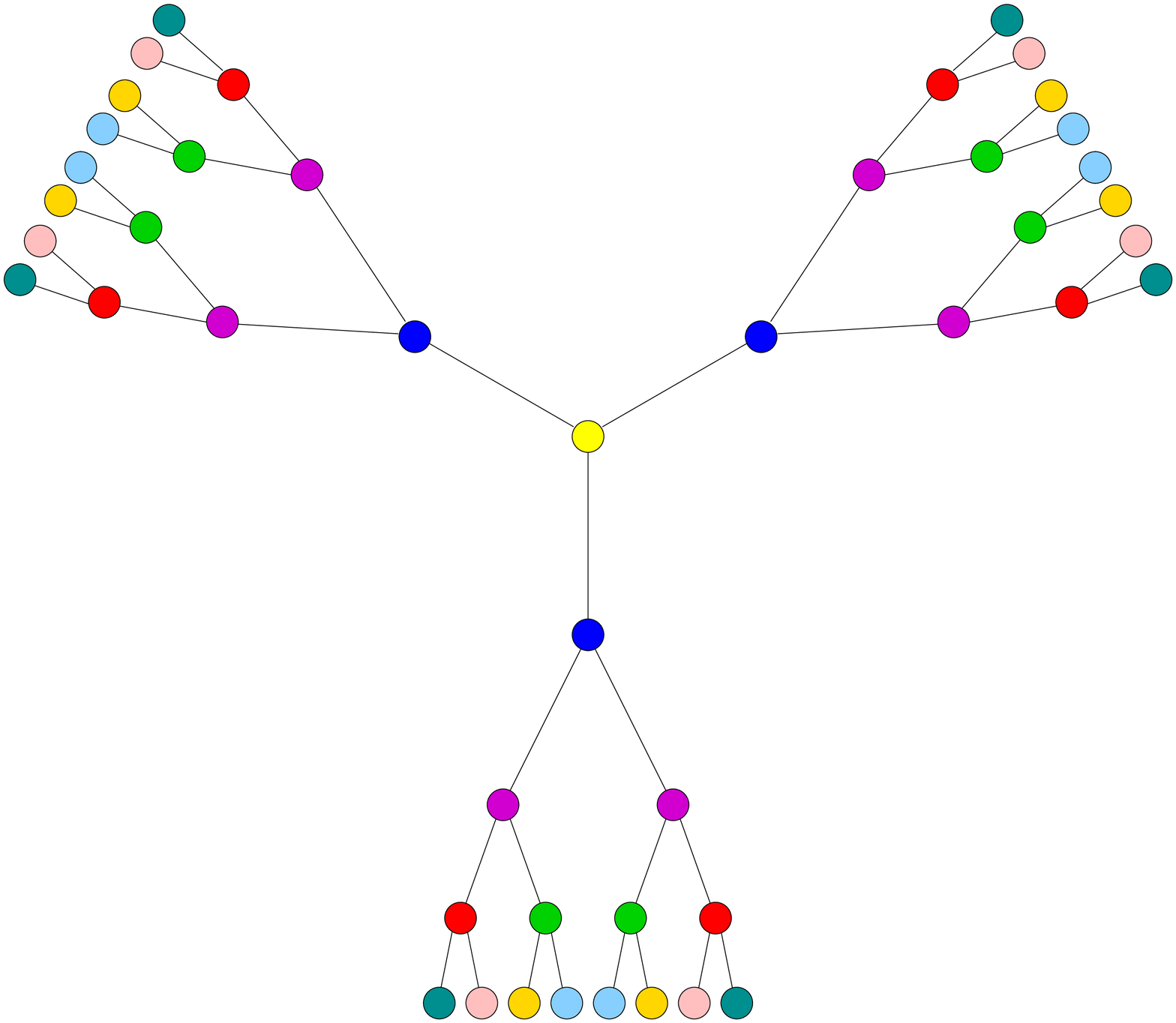}\\
\ea
\ea
\eeq

It is a beautiful fact that all del Pezzo surfaces give rise to
generalisations of the Markov equation \cite{exceptional}: the
exceptional collections on these surfaces are classified by such
equations. It was shewn in \cite{helix} that these equations coincide
with the associated Diophantine equations of the duality tree.

Of course, I am being cavalier with the name tree since in graph
theory a tree should not contain closed cycles. But I hope you will
indulge my botanic fancy. In fact, when the tree structure gets more
complicated, as we will see later, we will refer to the duality graph
as flowers.

%
\subsection{Cascades and Walls}
The last member of this trinity of dualities \cite{trio}
is the concept of {\bf
duality walls} coined in \cite{strass-wall}. This will perhaps be of
greater interest to the physicists in the audience. Thus far, we have
been dealing with conformal fixed points in the IR and the
beta-function vanishes. To be phenomenologically viable, we must allow
evolution of the renormalisation group. What we must do in the D-brane
probe picture is to introduce {\bf fractional branes}. Simply stated,
we must generalise the representation in \eref{Ni} to more
liberal choices of the labels $N_i$ so long as they satisfy the
anomaly cancellation conditions in \eref{anom}. Then, the
beta-function, out of conformality, can be thereby expressed using the
prescription of NSVZ \cite{nsvz}. The determination of the NSVZ
beta-function for quiver theories was performed in \cite{betafunc}.

How does the theory evolve with the beta-function? The answer was
supplanted in \cite{KS} for the simplest geometry, viz., the
conifold studied in \eref{conithy}. The beta functions for the two
nodes are of order $(M/N)^2$ where $M$ is the number of fractional
branes and $N$, the number of branes; the linear order, $\cO(M/N)$,
vanishes due to the $\IZ_2$ symmetry. We thus have
\beq\label{betacon}
\beta_1 = -3M+ {\cal O}(M/N)^2, 
\qquad \beta_2 = 3M + {\cal O}(M/N)^2.
\eeq
The two inverse gauge couplings evolve in linear fashion according to 
\eref{betacon} and when one reaches zero, i.e., the coupling becomes
infinite, we
perform Seiberg duality to map to a weak coupling regime. Hence
the inverse couplings $x_{i=1,2}=\frac{1}{g^2_{i}}$, when
plotted against $t = \log \mu$ of the energy scale $\mu$, evolve
in weave pattern, criss-crossing to infinite
energy. This is known as the {\it Klebanov-Strassler cascade} for the
conifold:
\beq\label{coni-cas}
\ba{cc}
\ba{l} \epsfxsize=4cm\epsfbox{quiver_conifold} \\
x_{i=1,2}=\frac{1}{g^2_{i}}, t = \log \mu \ea &
\ba{l}\epsfxsize=7cm\epsfbox{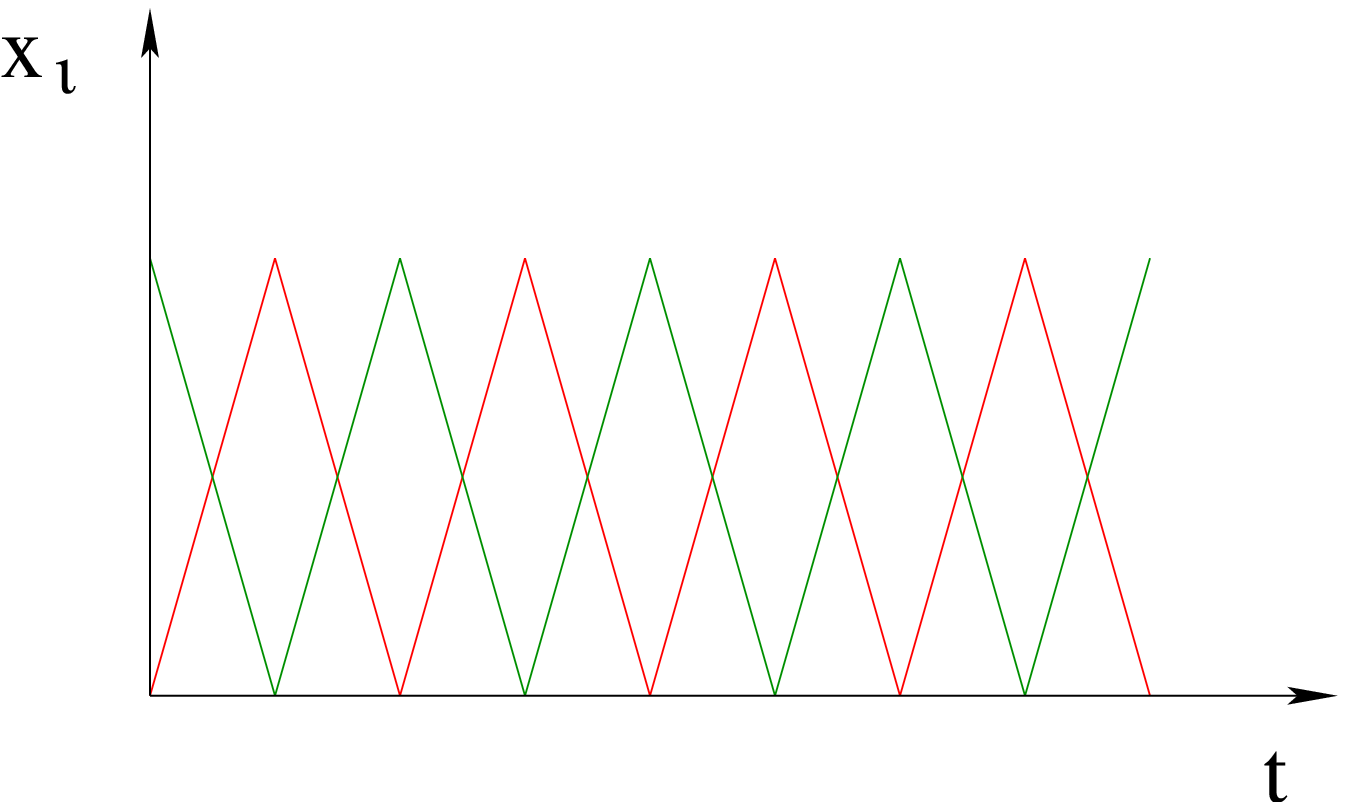}\ea
\ea
\eeq

The prescription is then evident if we wish to generalise this cascade
to other geometries as we will in the next subsection \footnote{
For a duality group perspective of general cascades, q.v.~\cite{halm}.
}.
We will remember the following rules:
(1) dualise whenever the
inverse coupling for the $i$-th node, $1/g_i^2 \rightarrow 0$;
(2) generically we obtain piece-wise linear $\beta_i$;
(3) for the $k$-th step in the dualisation, $1/g_i^2 \sim [\beta_i]_k
\Delta_k$ for step-size $\Delta_k$ in energy.
Conceptually, we are bouncing within simplicies in the space of
inverse gauge couplings.

%
\subsubsection{The Duality Wall and {\it Flos Hirzebruchiensis}}
A question was raised in \cite{strass-wall}, and subsequently answered
in the affirmative in \cite{fiol,HanWal} and \cite{walls1,chaos}, 
whether there
could ever be a {\bf duality wall} in a cascade for an arbitrary
singularity. In other words, could there be a case where the steps
$\Delta_k$ in energy scale during each dualisation decrease
consecutively, so that even after an infinite number of dualisations
one could not exceed a certain scale? This cutoff scale, where the
number of degrees of freedom in the gauge theory accumulates
exponentially, is called the duality wall.

Certainly, this phenomenon does not occur for the conifold: the steps
in the KS cascade
are constant in this case. However, if we took the next simplest case,
the cone over $F_0 = \IP^1 \times \IP^1$, 
the zeroth Hirzebruch surface, a markedly
different behaviour is noted.  
The quiver for this theory was presented in \eref{dPquiver}. We
reproduce it below, together with the first 11 dualisations, as well
as the duality tree:
{\vspace{-0.85cm}
\beq
\epsfxsize=7cm\epsfbox{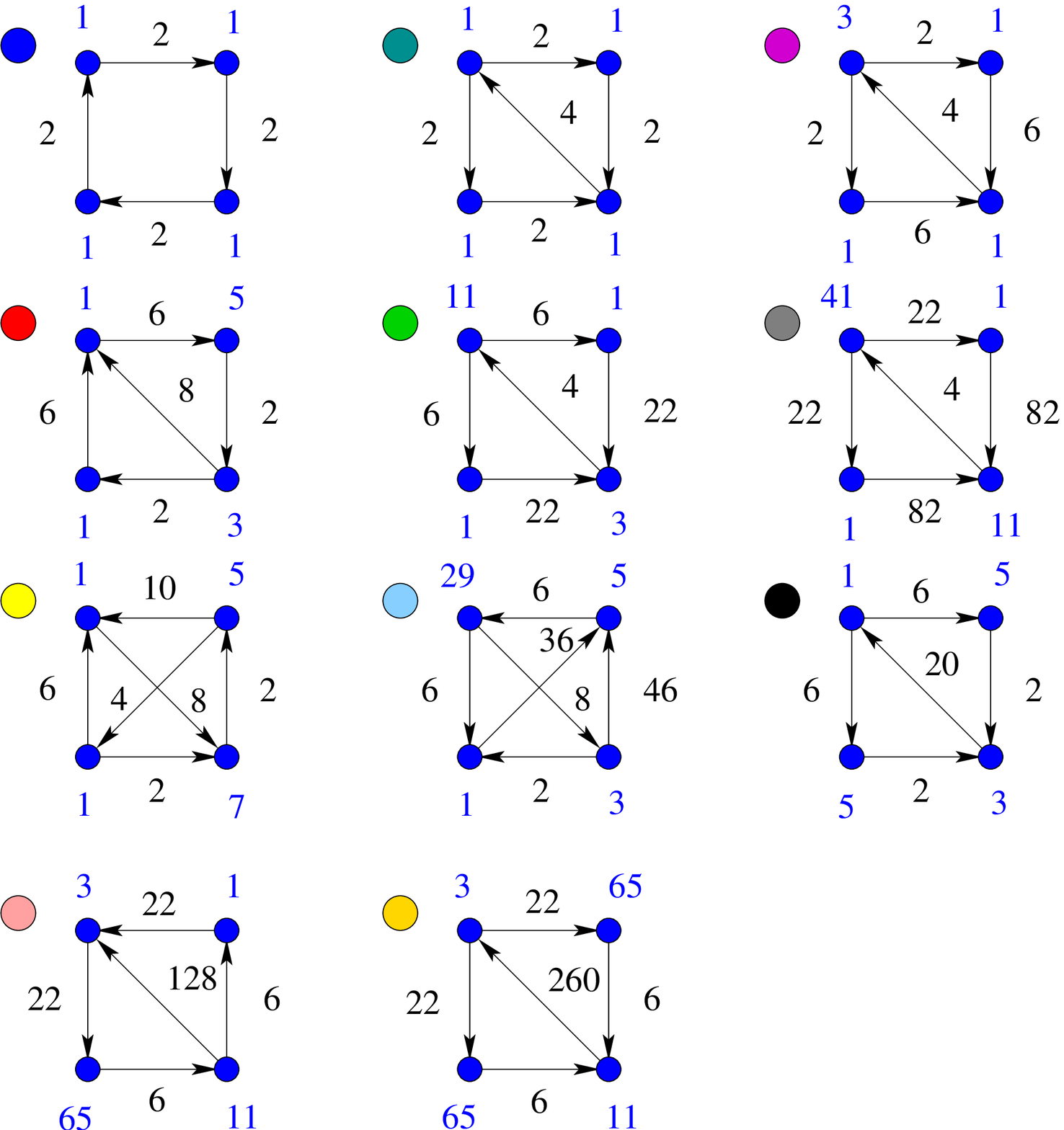} \quad
\epsfxsize=8cm\epsfbox{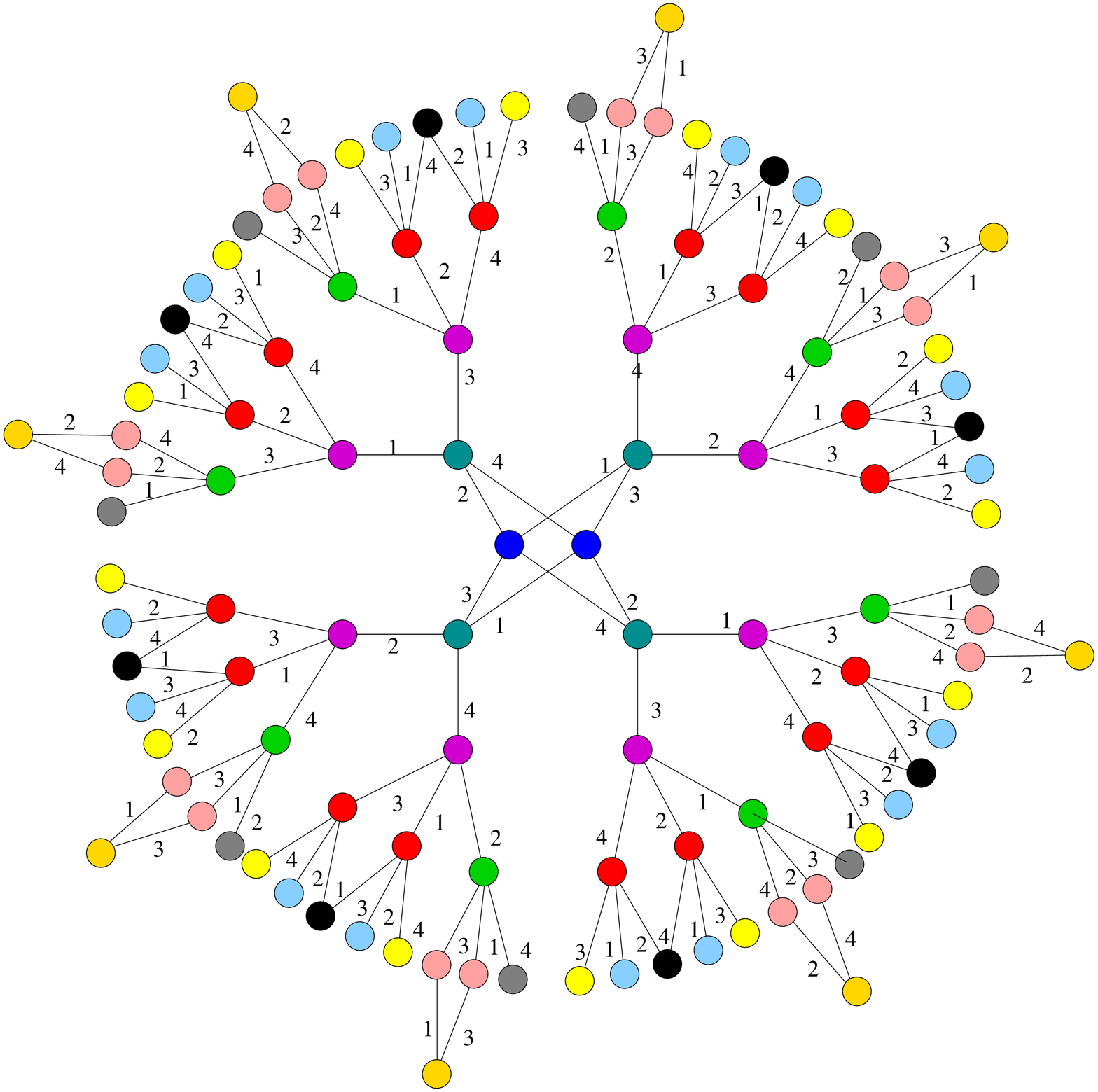}
\eeq}
Inspired by the aesthetic appeal of the duality structure, we could
not resist treasuring the result as a flower, 
which, with some affection, we call the {\it flos Hirzebruchiensis}.

Now, there are four nodes and hence the evolution of four gauge
couplings. If we dualised in the manner of 
alternating between the two phases
as shewn below, we would obtain the cascading behaviour
very much in the spirit of \eref{coni-cas}:
\beq
\ba{ll}
\ba{l}\epsfxsize=9cm\epsfbox{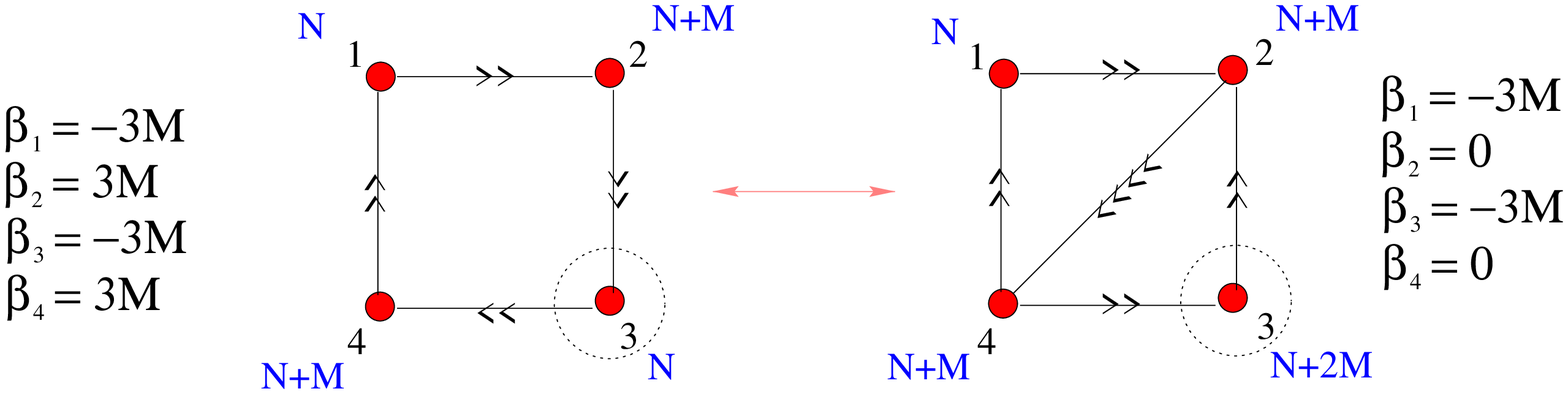}\ea &
\ba{l}\epsfxsize=7cm\epsfbox{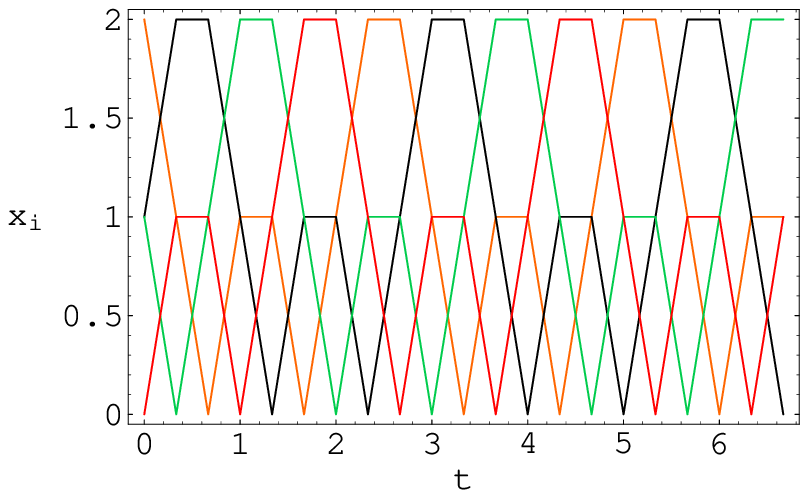}\ea
\ea
\eeq
However, if we dualised the phase below, we would observe an strikingly
apparent (and analytically proven \cite{walls1,chaos}) convergence to a wall.
\beq
\ba{ll}
\ba{l}\epsfxsize=6cm\epsfbox{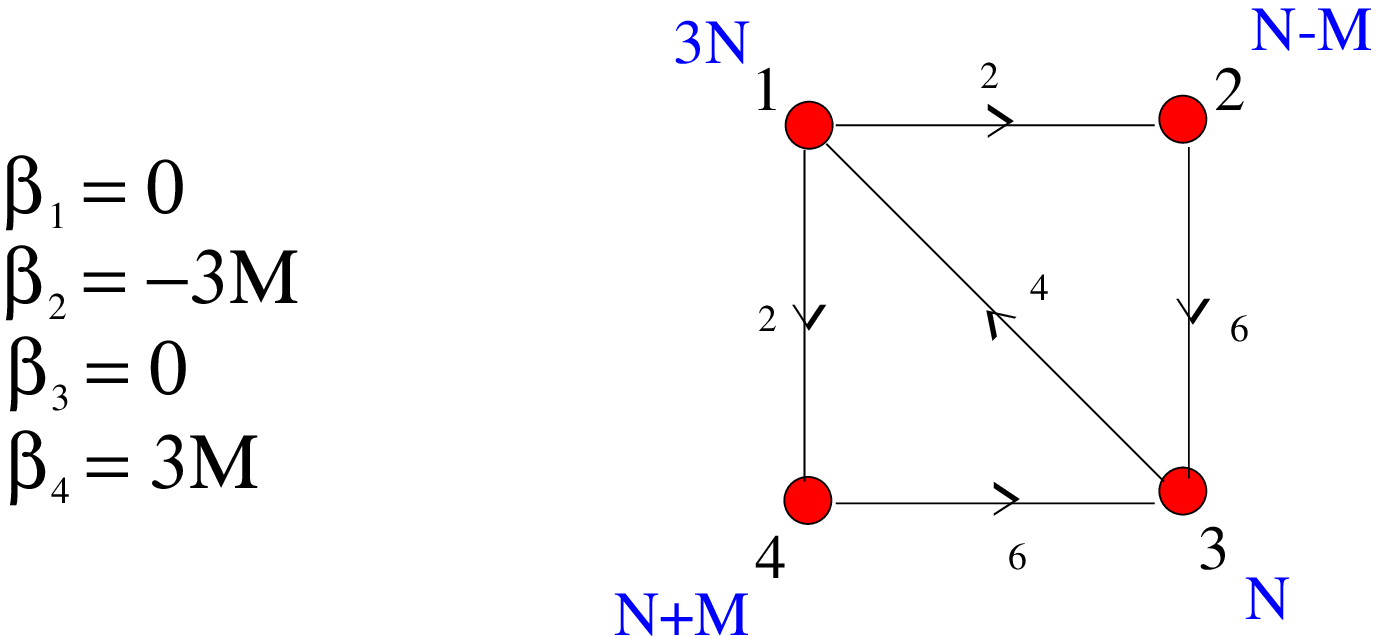}\ea&
\ba{l}\epsfxsize=6cm\epsfbox{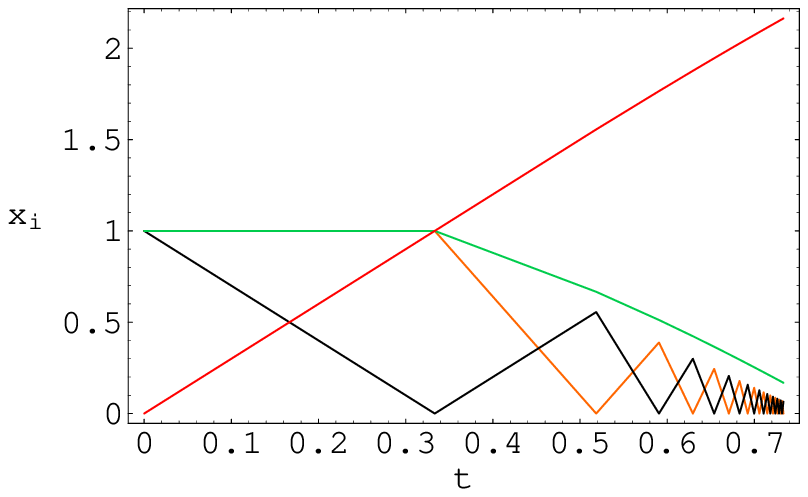}\ea
\ea
\eeq

We have mentioned above that the degree of freedom explodes at the
wall. This is indeed so. Recall from \eref{PL}
that the Seiberg duality action is a
monodromy matrix action $m$. At the $k$-step, as have an action $m^k$
on the labels $n_i$ of the initial quiver. On the other hand, 
the dof, i.e., the total amount of matter (arrows), is determined from
the labels via \eref{anom}. Therefore, the dof goes as the sum over
$\lambda^k$ where $\lambda$ are the eigenvalues of $m$. It was then
discussed in \cite{fiol} the conditions on $\lambda$ for which this
sum diverges; these quivers are known to the mathematicians as
hyperbolic.

The bloodhounds in the audiences may have acutely followed a scarlet
thread in this chromatic skein of our discussion. They have seen bouncing
in a simplex and the explosion of the degrees of freedom; do they not
smell chaos? It is indeed shewn in \cite{chaos} that the generic
geometry does give rise to cascades that exhibit chaotic behaviour. As
a tantalising last figure to our swift codetta, 
if one plotted the position of the wall
versus the initial gauge couplings (specifically, $t_{wall}$ against
$x_3(0)$ for $(1,1,x_3(0),0)$ for $F_0$), a self-replicating fractal
is seen:
\beq
\ba{l}\epsfxsize = 9cm\epsfbox{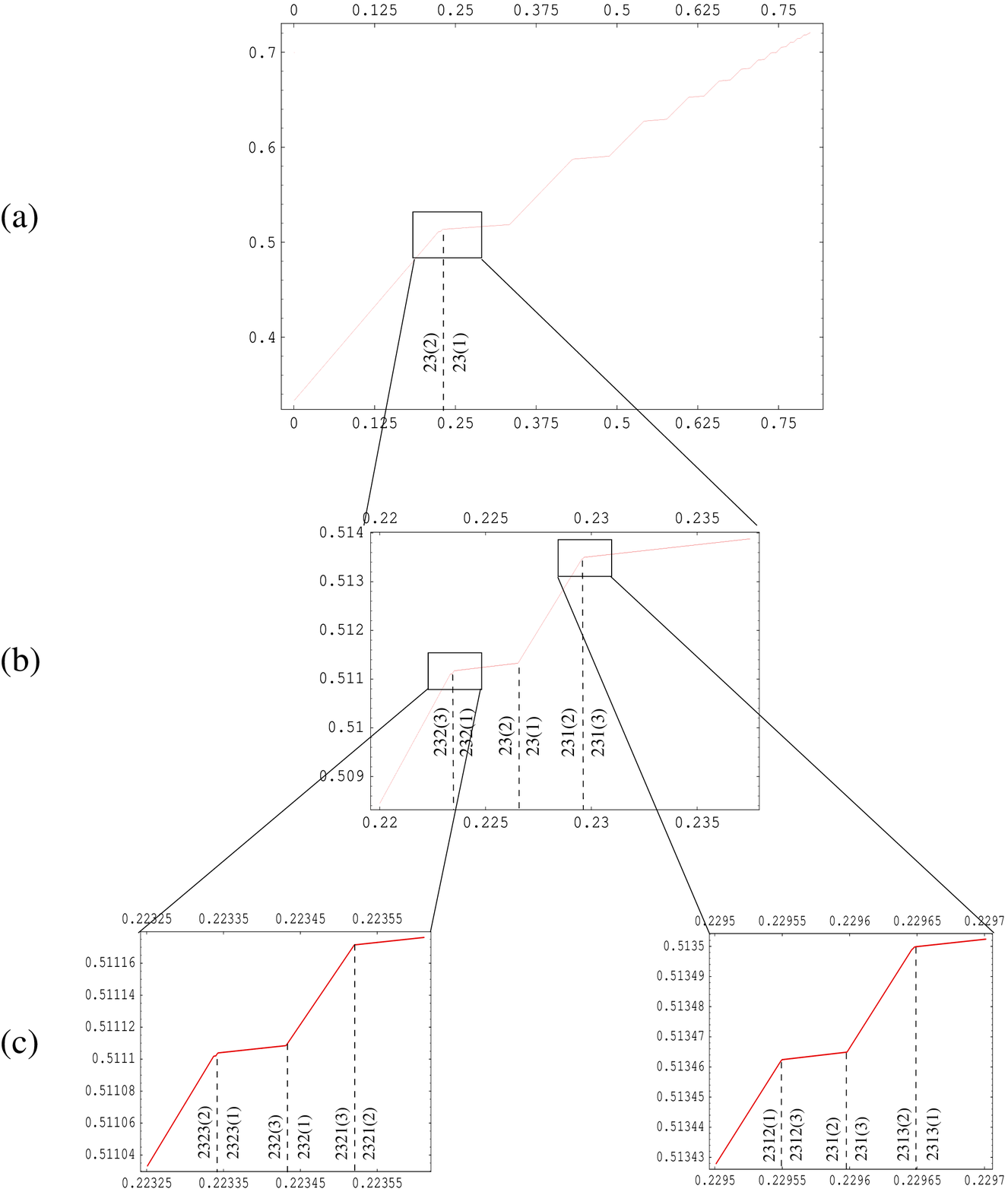}\ea
\eeq

\section*{Acknowledgements}
{\it Ad Catharinae Sanctae Alexandriae et Ad Majorem Dei Gloriam...\\}
I am honoured by and grateful to the organisers of the
HangZhou/BeiJing International Summer School in Mathematical Physics
for inviting me to give these lectures and am indebted especially to
Prof.~Chuan-Jie Zhu of the Chinese Academy of Sciences
for his hospitality as well as Prof.~Shing-Tung Yau, who has not only
done so much for mathematics but also so much for Chinese mathematics.

I would like to lend this opportunity to salute my brethren who have
toiled with me on the subject presented here: foremost, my former
advisor Amihay Hanany, whose non-chalant brilliance is a perpetual
inspiration and also my friends Bo Feng, Seba Franco, Chris
Herzog, Andreas Karch, Nikos Prezas, Jun Sung, Angel Uranga
and Johannes Walcher, who have cried
and laughed with me, the tears of scientific research.
Finally I thankfully acknowledge the gracious patronage of the
Dept.~of Physics and the Math/Physics RG of The University
Pennsylvania, a U.S.~DOE Grant $\#$DE-FG02-95ER40893 as well as
an NSF Focused Research Grant
DMS0139799 for ``The Geometry of Superstrings.''

\bibliographystyle{JHEP}

\end{document}